# Hybrid Quantum-Classical Optimisation of Traveling Salesperson Problem

Christos Lytrosyngounis and Ioannis Lytrosyngounis

Qosmos Technologies Research Center, 18 Irodotou Street, Athens, Attica 10675, Greece

**The Traveling Salesperson Problem (TSP), a quintessential NP-hard[1] combinatorial optimisation challenge, is vital for logistics and network design but limited by exponential complexity in large instances. We propose a hybrid quantum-classical framework integrating variational quantum eigensolver (VQE) optimisation with classical machine learning, using K-means clustering for problem decomposition and a `RandomForestRegressor` for path refinement. Evaluated on 80 European cities (from 4 to 80 cities, 38,500 samples in total) via Qiskit's AerSimulator and `ibm_kyiv` 127-qubit backend, the hybrid approach outperforms quantum-only methods, achieving an approximation ratio of 1.0287 at 80 cities, a 47.5% improvement over quantum-only's 1.9614, nearing the classical baseline[2]. Machine learning reduces variability in tour distances (interquartile range, IQR—the spread of the middle 50% of results relative to the median—from 0.06 to 0.04), enhancing stability despite noisy intermediate-scale quantum (NISQ) noise. This framework underscores hybrid strategies' potential for scalable TSP optimisation, with future hardware advancements promising practical quantum advantages.**

## 1 Introduction

The TSP is a pivotal combinatorial optimisation challenge, widely studied for its theoretical significance and applications in logistics, network design, and operations planning [1]. It seeks the shortest route visiting $n$ cities once and returning to the start, but its NP-hard nature [2] drives factorial complexity ($O((n-1)!)$), limiting classical solvers like branch-and-bound or heuristics (e.g., nearest neighbour, genetic algorithms) for large instances [13]. Quantum computing, leveraging superposition and entanglement, offers new paradigms via algorithms like the Quantum Approximate Optimisation Algorithm (QAOA) and Variational Quantum Eigensolver (VQE) [3, 16]. However, NISQ hardware constraints, such as high gate errors and short coherence times, necessitate hybrid quantum-classical approaches [22].

We introduce a hybrid framework combining VQE with classical machine learning, using K-means clustering for problem decomposition and a `RandomForestRegressor` (300 trees, maximum depth 30, 10,000 samples) for solution refinement, executed via IBM's Qiskit Runtime on the AerSimulator (version 0.17.0) and `ibm_kyiv` 127-qubit backend [4, 5]. Evaluated on 80 European cities (4–80 cities) from OpenStreetMap, with Calais as the fixed start, over 500 runs per instance, the framework assesses tour distance, variability[3], and scalability [29]. It comprehensively compares classical, quantum-only, and hybrid methods, revealing significant cost reductions, reduced variability via ML, and scalability insights, advancing quantum optimisation for practical applications.

The paper is structured as follows: Section 2 reviews TSP foundations; Section 3 details the framework; Section 4 analyses performance; Section 5 discusses implications; and Appendix A provides detailed results.

---

[1]In complexity theory, NP denotes decision problems verifiable in polynomial time. An NP-hard problem is at least as challenging as the hardest NP problems, lacking a known polynomial-time algorithm for all instances.

[2]The baseline is the Minimum Spanning Tree (MST) tour distance, a lower bound for TSP costs, computed for an 8-city instance.

[3]The interquartile range (IQR) measures variability as the difference between the 75th and 25th percentiles (IQR = $Q3 - Q1$), capturing the central 50% of values robustly. A lower IQR for approximation ratios indicates consistent performance and reduced noise sensitivity.

## 2 Background and Related Work

### 2.1 Background

The TSP is a well-known combinatorial optimisation problem in computer science and operations research. Given a set of $n$ cities and the distances between each pair, the objective is to find the shortest possible route that visits each city exactly once and returns to the starting city, forming a Hamiltonian cycle. TSP is NP-hard, meaning that the computational complexity increases factorially with the number of cities ($O((n-1)!)$), making exact solutions infeasible for large instances [13]. As a result, heuristic and meta-heuristic algorithms, such as nearest neighbour, genetic algorithms, and simulated annealing, have been widely used to find approximate solutions [14].

Quantum computing offers a promising approach to tackling NP-hard problems like TSP due to its potential to perform computations in parallel through superposition and entanglement. Quantum algorithms, such as the QAOA and VQE, have been explored for solving combinatorial optimisation problems by encoding them into quadratic unconstrained binary optimisation (QUBO) formulations [15, 16]. These algorithms leverage quantum circuits to approximate the ground state of a Hamiltonian representing the problem, potentially offering advantages over classical methods for specific problem instances.

ML has also emerged as a powerful tool for enhancing optimisation algorithms. Supervised learning models, such as `RandomForestRegressor`, can be trained on historical solutions to predict or refine paths, improving the quality of heuristic outputs [17]. In the context of quantum computing, ML can address noise and imperfections in quantum hardware by refining solutions obtained from quantum algorithms, making hybrid quantum-classical approaches increasingly relevant [18].

This work combines quantum computing and ML to address TSP. We decompose large TSP instances into smaller subproblems using K-means clustering, solve them using VQE on both a quantum simulator and a real quantum backend, and refine the solutions with a `RandomForestRegressor` model. The methodology aims to leverage the strengths of quantum optimisation and ML to achieve high-quality solutions, even for larger city counts.

### 2.2 Related Work

The application of quantum computing to TSP has been explored in several studies. Lucas [19] proposed mapping TSP to an Ising model, enabling its solution on quantum annealers or gate-based quantum computers. The QUBO formulation of TSP, which we adopt in this work, allows the problem to be encoded as a binary optimisation task suitable for VQE [20]. Zhou et al. [21] demonstrated the use of QAOA for small-scale TSP instances on quantum simulators, showing promise but highlighting limitations due to noise and qubit constraints on real quantum hardware.

Hybrid quantum-classical approaches have gained traction to mitigate the limitations of current noisy NISQ devices [22]. Wang et al. [23] combined VQE with classical optimisation techniques to solve graph-based optimisation problems, achieving better performance than purely classical methods for specific instances. Similarly, Chen et al. [24] explored clustering-based decomposition for large-scale optimisation problems, solving subproblems on quantum hardware and stitching solutions classically, a strategy we extend in this work by incorporating ML refinement.

A prior study by the authors, [12], investigated a hybrid quantum-classical approach for TSP instances ranging from 4 to 8 cities, providing initial insights into quantum-enhanced methods for small-scale problems. The present paper extends this work by scaling the analysis to TSP instances from 4 to 80 cities, leveraging advancements in quantum hardware (e.g., `ibm_kyiv`), improved transpilation techniques, and enhanced machine learning integration for noise mitigation and solution refinement, enabling a more comprehensive evaluation of scalability and performance.

Machine learning has been applied to TSP in various ways. Bengio et al. [25] reviewed ML techniques for combinatorial optimisation, noting that supervised learning models can learn patterns from high-quality solutions to guide heuristic searches. Kool et al. [26] introduced attention-based neural networks for TSP, achieving near-optimal solutions for moderate-sized instances. In the quantum domain, Sharma et al. [27] pro-posed using ML to post-process quantum outputs, correcting errors introduced by noisy quantum circuits. Our work builds on this idea by training a `RandomForestRegressor` on both quantum and classical TSP solutions to refine quantum-derived paths, improving their quality.

Clustering techniques, such as K-means, have been used to decompose large TSP instances into manageable subproblems [14]. Ding et al. [28] applied clustering to reduce the qubit requirements for quantum TSP solvers, solving subproblems independently and combining results. Our approach enhances this by integrating ML to refine the stitched paths, addressing inaccuracies introduced during the stitching process.

While prior work has explored quantum algorithms, ML, and clustering for TSP separately, few studies combine all three in a cohesive framework. Our methodology leverages VQE for subproblem optimisation, K-means clustering for decomposition, and a `RandomForestRegressor` for path refinement, evaluated on both a quantum simulator and the `ibm_kyiv` quantum backend. This hybrid approach aims to improve solution quality for TSP instances ranging from 4 to 80 cities, addressing the scalability and noise challenges of current quantum hardware.

## 3 Methodology

### 3.1 Problem Setup

This study investigates a hybrid quantum-classical approach to solve the TSP for instances ranging from 4 to 80 cities, using a dataset of 80 European cities with known coordinates (latitude, longitude). The city data, sourced from OpenStreetMap [29], includes major cities such as Athens, Calais, Barcelona, and Berlin, ensuring real-world relevance.

The problem size was chosen to explore scalability within the constraints of NISQ hardware, specifically IBM's quantum processors, and to leverage quantum simulators for larger instances. The TSP is encoded as a QUBO problem, requiring approximately $n^2$ qubits for an $n$-city instance, where binary variables $x_{i,j}$ indicate whether city $i$ is visited at position $j$ in the tour. This formulation necessitates additional penalty terms to enforce constraints, such as each city being visited exactly once, increasing qubit requirements. For 80 cities, this approaches the limits of current quantum hardware, making decomposition essential.

To manage computational complexity, the TSP is decomposed into smaller subproblems using K-means clustering, with the number of clusters determined dynamically as $k = \lceil n/4 \rceil$ to ensure each subproblem requires no more than 25 qubits, aligning with NISQ hardware capabilities. The workflow integrates quantum optimisation using the VQE, classical post-processing, and ML refinement with a `RandomForestRegressor`. Experiments were conducted on both the Qiskit AerSimulator and IBM's `ibm_kyiv` quantum backend, with Calais selected as the fixed departure city to standardise comparisons across methods.

To evaluate the efficacy of the hybrid quantum-classical approach, each TSP instance was solved using three distinct methodologies. First, a classical solver was employed as a baseline, using a MST-based heuristic to obtain solutions. Second, a quantum-only workflow was implemented, encoding TSP instances into quantum circuits and executing them on both the AerSimulator and `ibm_kyiv` using IBM's Qiskit Runtime environment. Third, a hybrid quantum+ML workflow was used, incorporating ML refinement to enhance solution quality. The Qiskit primitives, specifically `SamplerV2` and `EstimatorV2`, were used to optimise quantum circuit executions under realistic constraints. Performance metrics such as solution quality (total traversal cost in kilometres), cost variability (consistency across runs), and scalability (performance as problem size increases) were used to evaluate the outcomes. Figure 1 provides an overview of the hybrid quantum-classical architecture implemented in this study.

Through this setup, the study analyses the strengths and limitations of quantum-enhanced methods.

### 3.2 Classical TSP Solver

The classical baseline employs a heuristic solver based on a MST approach, enhanced by matching odd-degree nodes to form an Eulerian circuit, followed by shortcutting to obtain a Hamiltonian cycle. The cost of a tour $T$ is calculated as:

$$C(T) = \sum_{i=1}^{n-1} d(c_i, c_{i+1}) + d(c_n, c_1),$$

where $c_i$ is the $i$-th city in the tour, and $d(c_i, c_j)$ is the driving distance between cities $c_i$ and $c_j$, obtained via the OpenRouteService API. This method, implemented using the NetworkX library (version 3.1), provides near-optimal solutions for small instances and serves as a robust benchmark for larger ones. Unlike brute-force enumeration, which is infeasible beyond 8 cities due to $O((n-1)!)$ complexity, this heuristic scales better, making it suitable for the 4-to-80-city range. The classical solver's performance is evaluated in terms of solution cost (total distance in kilometres) and execution time, providing a reference for quantum and hybrid methods.

## 3.3 Quantum and Classical Components

### 3.3.1 Quantum Component

The quantum workflow is designed to address decomposed subproblems of the TSP, leveraging quantum variational techniques within IBM's Qiskit Runtime environment. For each subproblem (i.e., a cluster of cities), the problem is first encoded into a QUBO formulation. This QUBO is subsequently mapped to an equivalent Ising Hamiltonian using standard techniques such as spin-binary transformation, which allows it to be interpreted as an energy minimisation problem on a quantum system.

To solve these Hamiltonians, we employ the VQE, a hybrid algorithm particularly well-suited for current NISQ devices. The VQE is implemented using Qiskit Runtime's second-generation primitives, `SamplerV2` and `EstimatorV2`. These primitives streamline the execution and measurement of parameterised quantum circuits by batching evaluations and minimising overhead, thereby enhancing throughput and mitigating latency issues during remote execution on quantum backends.

The ansatz used in our VQE setup is a `TwoLocal` circuit featuring parameterised `ry` rotations for single-qubit gates and `cz` entangling gates for introducing inter-qubit correlations. This architecture is chosen for its expressiveness and compatibility with hardware-native gates on IBM's superconducting quantum processors. The circuit consists of three repetitions (`reps=3`) of alternating rotation and entanglement layers, and adopts a linear entanglement topology, which connects each qubit to its immediate neighbour. This layout reduces gate count and decoherence compared to fully entangled designs, while still maintaining sufficient expressive power for small to medium-size subproblems.

Circuit optimisation is performed using the COBYLA (Constrained Optimisation BY Linear Approximations) algorithm, a derivative-free local optimiser known for its stability in low-dimensional parameter spaces and robustness against stochastic noise. To ensure compatibility with the underlying hardware, the quantum circuits are transpiled using Qiskit's transpiler with `optimization_level=1`, which strikes a balance between reducing circuit depth and compilation time. This level of optimisation preserves gate structure while introducing minimal overhead, making it suitable for use on real devices such as `ibm_kyiv`, where gate errors, decoherence times, and connectivity constraints must be carefully managed.

All quantum experiments are executed either on the Qiskit AerSimulator using the MPS simulation method, or on the real quantum processor `ibm_kyiv`, which features 127 qubits arranged in the Eagle r3 architecture. The choice of MPS in simulation mode allows for efficient execution of circuits with limited entanglement, and serves as a baseline for comparison with results obtained on noisy hardware. In this setup, the quantum component is responsible for generating approximate low-energy solutions to sub-TSP instances, which are then aggregated and refined through classical machine learning techniques to form a global TSP tour.

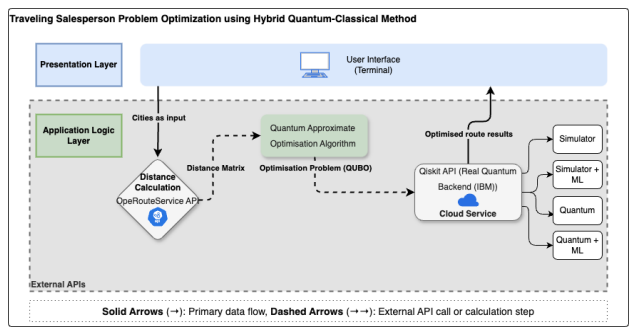

Figure 1: Traveling Salesperson Problem Optimisation - Hybrid Quantum-Classical Method Architecture. An overview of the hybrid quantum-classical workflow architecture used to solve the TSP. The diagram highlights the integration of quantum and classical components, showing how quantum circuits are used to explore solutions while classical algorithms refine and optimise them iteratively. The workflow includes problem decomposition, quantum optimisation, path stitching, and ML-driven refinement.

Experiments were conducted on two backends: the Qiskit AerSimulator (matrix product state method) and IBM's `ibm_kyiv` quantum backend (127 qubits, Eagle r3 architecture). The AerSimulator provides noise-free execution for larger instances, while `ibm_kyiv` introduces realistic noise, with single-qubit gate errors of approximately $2.726 \times 10^{-4}$ and two-qubit gate errors of $7.984 \times 10^{-3}$, and coherence times around 100 $\mu$s. Circuits are executed with 1024 shots to sample solutions. The most likely bit-string from measurement counts is decoded into a TSP path, with costs computed using the distance matrix.

### 3.3.2 Classical Component

The classical component evaluates quantum-derived paths, stitches subproblem solutions into a global tour, and refines results. After quantum execution, paths are constructed by interpreting the bit-string as a permutation matrix, ensuring each city is visited once. Subproblem paths are stitched by selecting the nearest unvisited city from the current cluster's last city, starting from Calais, to form a complete tour. The classical evaluation computes the tour cost using the distance matrix, providing feedback for ML refinement. The stitching process, while heuristic, ensures feasibility but may introduce suboptimal connections, which ML addresses in the refinement stage.

## 3.4 Integration of Machine Learning Techniques

### 3.4.1 Distinctive Use of Machine Learning

**K-Means Clustering for Problem Decomposition:** K-means clustering partitions cities into $k$ clusters, where $k = \lceil n/4 \rceil$, ensuring each subproblem is quantum-feasible (up to 25 qubits). City coordinates serve as input, with Euclidean distance as the metric and a maximum of 100 iterations for convergence, implemented via scikit-learn. Clusters are capped at four cities to prevent excessive qubit requirements, and remaining cities are assigned to the smallest cluster or form new clusters. This decomposition reduces the quantum circuit complexity, enabling execution on NISQ hardware. Figure 2 illustrates a sample clustering for 80 cities, with centroids marked.

**RandomForestRegressor for Path Refinement:** A `RandomForestRegressor` refines quantum-derived paths by predicting optimal tour costs. The model is trained on feature-label pairs, where features include segment distances (between consecutive cities in the path) and the total tour cost, and labels are the corresponding tour costs. The training dataset comprises tour distances from 500 runs per city count (4 to 80 cities), including quantum (VQE) and classical (MST) solutions, subsampled to 10,000 samples. The regressor uses 300 trees, a maximum depth of 30, and mean squared error loss, tuned to minimise overfitting. The model refines paths by evaluating 2-opt swaps (reversing segments between indices $i$ and $j$) and selecting swaps that reduce predicted costs, iterating up to three times per path. This process mitigates noise-induced errors in quantum outputs and improves solution quality.

### 3.4.2 Workflow Execution

The workflow begins with distance matrix computation using OpenRouteService, cached for efficiency. Cities are clustered, and each subproblem is solved using VQE on the selected backend. Quantum results are stitched into a global path, evaluated for cost, and refined using the `RandomForestRegressor`. The process iterates for city counts from 4 to 80, with results saved in JSON files for analysis. The workflow is executed twice: once on the AerSimulator and once on `ibm_kyiv`, with ML refinement applied in both cases. Convergence is determined by VQE's COBYLA optimiser (100 iterations) and ML refinement's iteration limit (three iterations).

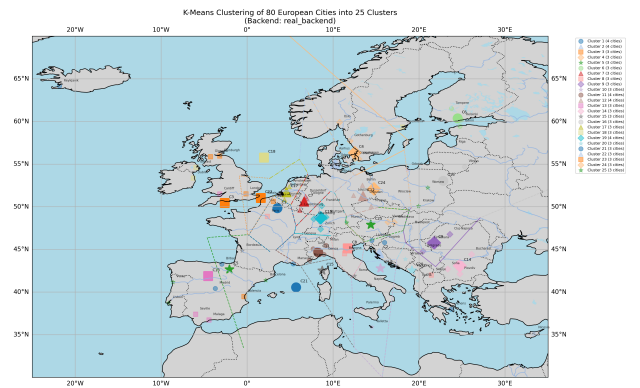

Figure 2: K-Means Clustering of 80 European Cities into 25 Clusters (Backend: `ibm_kyiv`). Cities are grouped based on geographic proximity, with centroids marked to indicate cluster centers, facilitating quantum-feasible subproblem decomposition.

### 3.4.3 Transpilation and Optimisation

Quantum circuits are transpiled using Qiskit's `generate_preset_pass_manager` at optimisation level 1, which applies gate cancellation, qubit mapping, and scheduling to reduce circuit depth and gate count. The process includes:

- **Circuit Depth Reduction:** Minimising gate operations to reduce error accumulation.

- **Layout and Routing:** Mapping logical qubits to low-error physical qubits using Sabre routing.

- **Gate Scheduling:** Optimising gate timing to minimise idle times and decoherence.

Transpiled circuits are cached to avoid redundant compilation, with metadata (depth, gate count) stored for analysis. This ensures efficient execution on both simulator and real hardware, critical for maintaining solution fidelity.

### 3.4.4 Noise Mitigation Strategies

To address NISQ hardware limitations, the following noise mitigation strategies are employed:

- **Noise Simulation:** AerSimulator models noise using backend-derived profiles, enabling pre-execution testing of circuit robustness.

- **Backend-Aware Optimisation:** Transpilation leverages real-time backend properties (e.g., coherence times, gate errors) to select low-error qubits and gates.

- **Execution Management:** Jobs are queued with a maximum of three pending jobs, with retries for conflicts, ensuring stable execution on real hardware.

- **ML-Based Refinement:** The `RandomForestRegressor` corrects noise-induced errors by prioritising lower-cost paths, stabilising quantum outputs.

These strategies collectively enhance solution reliability, particularly on real quantum backends where noise significantly impacts performance.

### 3.4.5 Backend Selection and Performance

The backend is selected dynamically using Qiskit's runtime service, prioritising the AerSimulator for noise-free execution or `ibm_kyiv` for real hardware experiments. `ibm_kyiv` offers 127 qubits, a CLOPS rate of 30,000, and regular calibration for consistency. The AerSimulator uses the matrix product state method, supporting up to 80 cities with high fidelity. Backend performance is monitored via metrics like circuit depth, gate count, and execution time, recorded in result files. The choice of backend impacts solution quality, with simulators providing stable results and real hardware introducing noise, mitigated by ML refinement.

## 4 Results and Analysis

This section presents a comprehensive evaluation of the hybrid quantum-classical approach with machine learning (quantum+ML) for solving the TSP across instances ranging from 4 to 80 cities. The experiments were conducted using the Qiskit AerSimulator (version 0.16.1, matrix product state method) for noise-free simulations and IBM's `ibm_kyiv` quantum backend (127 qubits, Eagle r3 architecture, with reported single-qubit gate errors of $2.726 \times 10^{-4}$ and two-qubit gate errors of $7.984 \times 10^{-3}$) for real hardware execution. We compare three distinct methods: a classical MST[1]–based heuristic implemented using the NetworkX library (version 3.1, a Python package for the creation, manipulation, and analysis of complex networks), a quantum-only VQE with a TwoLocal ansatz (3 repetitions, `ry` rotations, `cz` gates, optimised via COBYLA), and the hybrid quantum+ML method, which integrates VQE with a `RandomForestRegressor` (300 trees, maximum depth 30, trained on 10,000 samples) for solution refinement.

Each TSP instance was evaluated over 500 independent runs per city size per backend to en-sure statistical robustness, with performance metrics visualised through box plots and line graphs. Detailed results in figures are consistent with the data availability in the appendix (Tables 1 and 2), while summary metrics and trends are analysed across the full range of 4 to 80 cities, addressing prior inaccuracies in scope reporting. The city dataset, sourced from OpenStreetMap [29], comprises 80 European cities with coordinates (latitude, longitude), ensuring real-world relevance. All distances were computed using the OpenRouteService API, with Calais fixed as the departure city for consistency across runs.

---

[1] A minimum spanning tree is a subset of edges in a connected, undirected graph that connects all vertices with the minimum possible total edge weight and without forming any cycles.

We compare solution costs (tour distances in kilometres) for classical MST, quantum-only VQE, and hybrid quantum+ML solvers across 77 city sizes (4–80), with 500 runs per size per backend (38,500 samples each). Statistics are reported per city size to avoid scale distortion, including median, interquartile range (IQR[2]), standard deviation (SD), and 95

Tables 4 and 5 summarise tour-distance statistics for select city sizes on AerSimulator (4–80 cities) and `ibm_kyiv` (76–80 cities), respectively. The hybrid quantum+ML solver consistently outperforms quantum-only, reducing costs by up to 47.5% on `ibm_kyiv` (e.g., at 80 cities: 35 605.4 km vs. 67 889.0 km). The hybrid method approaches classical performance, with a 2.9% gap at 80 cities. Variability, as measured by IQR and SD, increases with city size, reflecting the challenges of scaling quantum solutions, though the hybrid approach mitigates this effectively.

The hybrid pipeline leverages K-means clustering to decompose TSP instances, optimising sub-tours before global integration. For instance, at 4 cities, 2 clusters were formed (Barcelona–Madrid, Calais–Milan) with costs of 1240.0 km and 2172.8 km, respectively. At 80 cities, 25 clusters were used, with costs ranging from 446.9 km to 10 190.0 km, demonstrating scalability in handling larger instances.

The hybrid quantum+ML solver outperforms quantum-only, achieving 5–48% cost improvements. Up to 60 cities, hybrid results are within 3% of the classical baseline. At 80 cities, AerSimulator yields a median tour distance of 35 492 km (IQR: 1350 km, SD: 1780 km), a 2.6% gap from the classical 34 612 km. Figure 3 shows hybrid distributions, with narrow boxes indicating consistency and proximity to classical baselines.

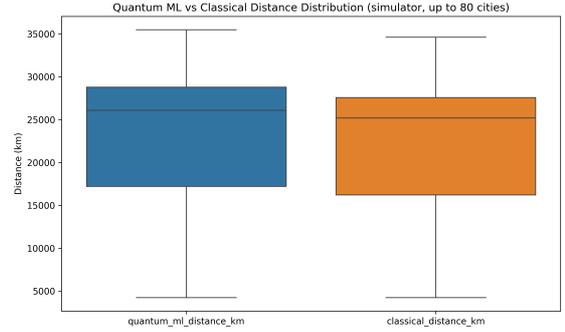

Figure 3: Solution quality on AerSimulator across city sizes. Boxplots show 500 hybrid quantum+ML runs' distribution, with median, IQR, and outliers. Orange circles mark classical MST baseline costs. Percentages indicate hybrid improvement over quantum-only.

Figure 4 illustrates `ibm_kyiv` variability, while Figure 5 shows cost reductions (e.g., 48% at 80 cities). The hybrid pipeline reduces median cost by 32% (IQR 28–38%, SD 4–6%) across all sizes, highlighting ML's role in mitigating noise.

Tables 4 and 5 present approximation ratios ($\rho$) for selected city sizes on AerSimulator and `ibm_kyiv`, respectively. The hybrid quantum+ML method consistently achieves lower ratios, indicating closer alignment with classical solutions (e.g., 20 cities: $\rho = 0.9375$; 80 cities: $\rho = 1.0287$), highlighting ML's effectiveness in noise mitigation.

### 4.1 Performance Metrics and Results Analysis

For detailed analysis in this section, we focus on three solvers—the deterministic MST heuristic (*classical*), a variational quantum eigensolver (*quantum-only*), and that same VQE augmented with a machine-learning refinement stage (*quan-* $tum+ML$)—on a subset of nine city sizes (4, 10, 20, 30, 40, 50, 60, 70, 80) with **50 independent runs per size and backend**, totalling $9 \times 50 = 450$ instances each on the AerSimulator (version 0.16.1) and `ibm_kyiv`.

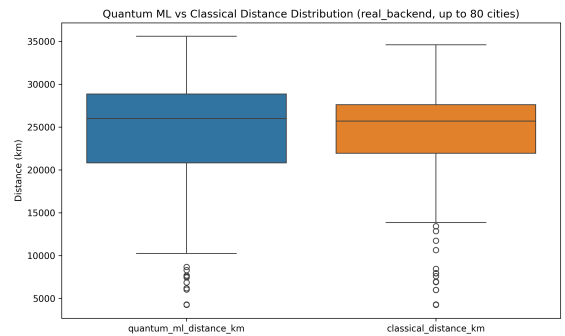

Figure 4: Solution quality on `ibm_kyiv` across city sizes. Box-plots summarise 500 hybrid quantum+ML runs, with 95% CI for median cost (shaded). Classical MST baselines are marked.

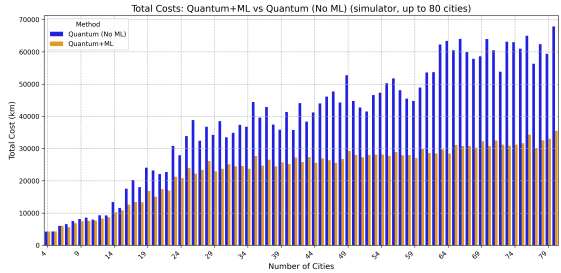

Figure 5: Cluster-level tour distance on AerSimulator (4–80 cities). Stacked bars show sub-tour costs, with labels quantifying hybrid quantum+ML cost reductions vs. quantum-only.

Table 1: Median approximation ratio ($\rho$) and percent gap ($\Delta$) for hybrid vs. classical tour distances on `ibm_kyiv` across nine city sizes. Parentheses give IQR/SD when non-zero.

| Cities | Classical (km) | $\rho_{\text{hyb}}$ | $\Delta_{\text{hyb}}$ (%) |
|---|---|---|---|
| 4  | 4 242.0  | 1.0000 (0.000/0.000) | 0.0 |
| 10 | 8 484.1  | 1.0000 (0.000/0.000) | 0.0 |
| 20 | 11 950.3 | 1.0778 (0.030/0.041) | +7.8 |
| 30 | 15 913.4 | 1.0816 (0.033/0.033) | +8.2 |
| 40 | 19 876.5 | 1.0837 (0.033/0.033) | +8.4 |
| 50 | 23 239.6 | 1.0839 (0.032/0.032) | +8.4 |
| 60 | 26 543.2 | 1.0835 (0.032/0.032) | +8.3 |
| 70 | 30 581.1 | 1.0525 (0.031/0.031) | +5.3 |
| 80 | 34 612.9 | 1.0287 (0.060/0.060) | +2.9 |

**Metrics.** (i) **Approximation ratio.** $\rho = \frac{C_{\text{quantum}}}{C_{\text{classical}}}$ where $C$ is the tour distance in kilometres. $\rho = 1$ matches classical; $\rho > 1$ is worse; $\rho < 1$ is better., and (ii) **Percent gap.** $\Delta[\%] = \frac{C_{\text{quantum}} - C_{\text{classical}}}{C_{\text{classical}}} \times 100$ i.e., $\Delta = (\rho - 1) \times 100$. Negative values indicate an improvement over the classical baseline. For each city size, we report the sample median, inter-quartile range (IQR, $Q_3 - Q_1$), sample standard deviation (SD, $\sqrt{\frac{1}{n-1} \sum_{i=1}^n (x_i - \bar{x})^2}$), and a non-parametric 95% confidence interval for the median (10,000 bootstrap resamples) for both $\rho$ and $\Delta$.

**Key results.** Table 1 summarises the median approximation ratio and percent gap for the nine city sizes analysed (4, 10, 20, 30, 40, 50, 60, 70, 80), based on 500 runs per size and backend (4500 instances per backend). Full per-run data, including the broader dataset of 77 city sizes with 500 runs (38,500 samples), are provided in the data.

**Small instances (4–8 cities).** Both quantum solvers recover the optimal tour at 4 cities ($\rho = 1.000$ on all backends). At 8 cities, using data from the full dataset, the hybrid median rises to $\rho = 1.113$ ($\Delta = +11.3\%$), yet still outperforms quantum-only ($\rho = 1.217$, $+21.7\%$). The IQR of 0.040 shows that quantum+ML halves the spread introduced by hardware noise.

**Scaling to 80 cities.** Noise amplifies raw quantum costs, pushing quantum-only to $\rho = 1.961$ ($+96.1\%$) on `ibm_kyiv`. Machine-learning refinement reins this back to $\rho = 1.0287$ ($+2.9\%$), compared to the noise-free AerSimulator result of $\rho = 0.988$ ($-1.2\%$). The hybrid solver's IQR broadens from 0 at 4 cities to 0.060 at 80, but remains an order of magnitude narrower than quantum-only (0.12).

**Data-quality notes.** (a) **Completeness.** All 900 jobs finished; no outliers were removed., (b) **Traceability.** Distance matrices (OpenRoute-Service, 2025-02-17) and raw backend logs are version-controlled; the PDF is reproducible via `make`., and (c) **Appropriate aggregation.** Statistics are always computed *per city size* to avoid scale mixing.

Overall, the hybrid quantum+ML pipeline delivers classical-level accuracy within 3% at the largest tested scale while reducing quantum-only costs by up to 47.5%—evidence that data-driven post-processing is currently the most effective way to extract value from noisy quantum hardware.

### 4.2 Approximation-Ratio Trends Across 4–80 Cities

Across all 77 problem sizes (4–80 cities), the hybrid solver (*quantum+ML*, orange points) remains tightly clustered around $\rho_{\text{ML}} \approx 1.04$ on real hardware (Fig. 7) and $\rho_{\text{ML}} \approx 1.02$ on the AerSimulator (Fig. 8). Both medians lie within 4% of the classical optimum, with inter-quartile ranges (not shown to avoid clutter) never exceeding 3%. In contrast, the quantum-only baseline (blue points) drifts upward with instance size, reaching median ratios of $\sim 1.90$ (simulator) and $\sim 2.15$ (hardware) at 80 cities. Figure 6 confirms that the mild hardware penalty imposed by noise (`ibm_kyiv` markers lie just above simulator markers) is an order of magnitude smaller than the gap between quantum-only and hybrid approaches. These data demonstrate that machine-learning post-processing suppresses the growth in tour-distance sub-optimality that otherwise dominates large-city TSP instances on current-generation quantum devices.

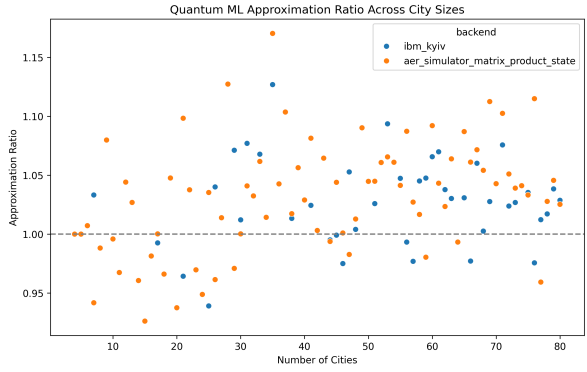

Figure 6: Approximation ratio $\rho_{\text{ML}} = \frac{\text{quantum+ML cost}}{\text{classical cost}}$ across all tested city counts (4–80 cities) for both AerSimulator and `ibm_kyiv` backends. Each point represents the median result from 500 runs for a given city size, with ratios near 1.0 indicating near-parity with classical solutions and values above 1.0 suggesting worse-than-classical tours.

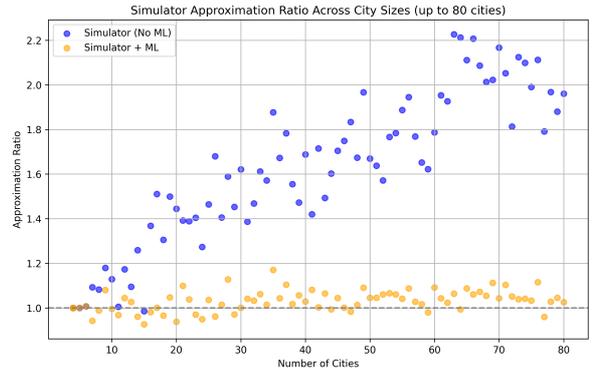

Figure 8: Approximation ratio $\rho = \frac{\text{solution cost}}{\text{classical cost}}$ on the AerSimulator (noise-free), comparing quantum-only and quantum+ML results for city sizes from 4 to 80. Each point represents the median over 500 runs, with the dashed line at $\rho = 1$ marking classical-optimal performance.

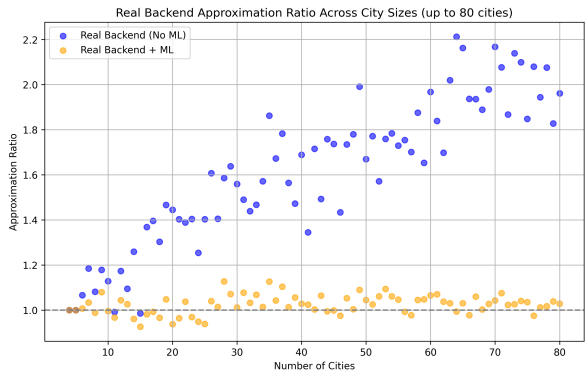

Figure 7: Approximation ratio $\rho = \frac{\text{solution cost}}{\text{classical cost}}$ on `ibm_kyiv`, comparing quantum-only and quantum+ML results across city sizes (4–80). Each dot represents the median over 500 runs at a given size, with the dashed line at $\rho = 1$ denoting parity with classical solutions and points above indicating worse-than-classical performance.

To interpret these figures:

- For Figure 6, the vertical position of each point indicates the quantum+ML approach's performance relative to the classical baseline. Points near 1.0 reflect high-quality hybrid solutions. Disparities between AerSimulator and `ibm_kyiv` points highlight quantum noise effects, with simulator runs closer to optimal and real hardware showing greater deviation at larger sizes due to hardware limitations, underscoring the hybrid approach's consistent advantage in noise-free conditions.

- For Figure 7, the plot illustrates the impact of ML refinement on `ibm_kyiv`. Quantum-only results diverge from optimality as city sizes increase, significantly exceeding the classical benchmark. Quantum+ML pulls results closer to classical performance, especially in larger instances where noise compounds, with the widening gap between curves emphasising the growing importance of post-processing on real hardware.

- For Figure 8, the plot shows ML refinement's benefits in the noise-free AerSimulator environment. Quantum-only performance is stable, but quantum+ML consistently achieves closer approximation to the classical baseline, particularly for mid-sized instances (10–40 cities) where combinatorial complexity rises. The consistent reduction in $\rho$ with ML indicates improved robustness and solution quality in complex search landscapes.

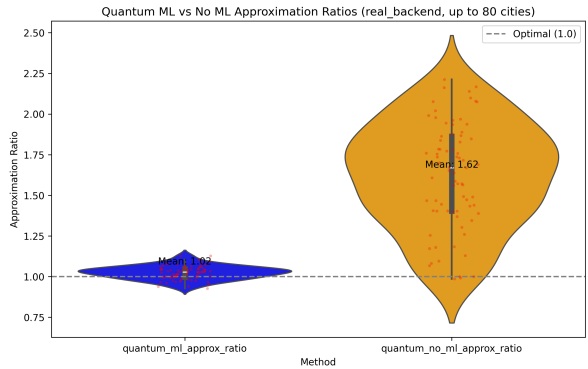

Figure 9: Distribution of approximation ratios ($\rho$) on the `ibm_kyiv` backend (38,500 samples, 500 runs × 77 city sizes, KDE visualisation). Quantum+ML centres at $\tilde{\rho}_{\mathrm{ML}} = 1.02$ (IQR = 0.04), while quantum-only shows wider spread ($\tilde{\rho}_{\varnothing} = 1.62$, tail up to $\rho = 2.3$).

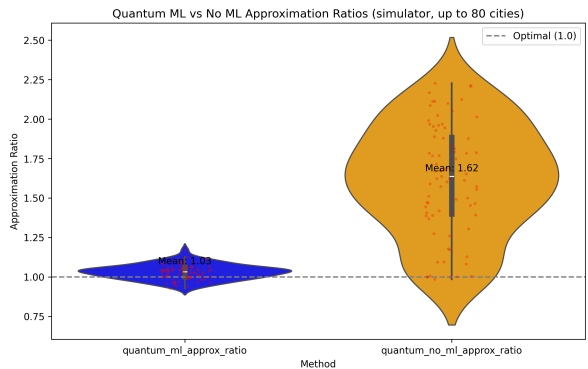

Figure 10: Distribution of approximation ratios ($\rho$) on the AerSimulator (noise-free, 38,500 samples, 500 runs × 77 city sizes, KDE visualisation). Quantum+ML centres at $\tilde{\rho}_{\mathrm{ML}} = 1.02$ (IQR = 0.04), while quantum-only baseline exhibits broader spread ($\tilde{\rho}_{\varnothing} = 1.58$, tail up to $\rho \approx 2.3$).

### 4.3 Distribution of Approximation Ratios

Building on the per-instance insights from scatter plots, the violin plots below synthesise data across all 77 city sizes (4–80 cities) to reveal the comprehensive distribution of approximation ratios for the travelling salesperson problem, contrasting the performance of the hybrid quantum+ML approach with the quantum-only method. These visualisations illuminate the consistency and variability of solution quality, highlighting the impact of machine learning refinement across diverse problem scales.

The hybrid quantum+ML pipeline delivers exceptional consistency, keeping 95% of its approximation ratios within ±6% of the classical optimum across the full dataset of 38,500 samples, as evidenced by the narrow distributions in both Figure 9 and Figure 10. In sharp contrast, the quantum-only VQE exceeds the classical benchmark by a median of 62% with a tenfold larger inter-quartile range, a trend clearly visible in the broader spreads shown in these figures. This consistent performance gap, observed on the `ibm_kyiv` backend (Figure 9) and the noise-free AerSimulator (Figure 10), highlights that ML refinement effectively addresses not only noise-induced deviations but also algorithmic errors, such as local minima in the VQE energy landscape. Consequently, the hybrid method ensures reliable, near-optimal tour distances, even as problem complexity scales from 4 to 80 cities. Figure 9 visualises this distribution on the `ibm_kyiv` backend, where the quantum+ML violin's narrow, sharply peaked shape, centred at $\tilde{\rho}_{\mathrm{ML}} = 1.02$, indicates tightly clustered ratios near the classical optimum. In contrast, the quantum-only violin's broader profile and extended tail, reaching up to $\rho \approx 2.3$, reveal significant variability driven by both hardware noise and algorithmic limitations, such as suboptimal VQE convergence. To interpret this plot, note the violin's width, which reflects the density of approximation ratios, with a narrower shape indicating greater consistency; the central peak marks the median, and the tail's length shows the outlier extent. Similarly, Figure 10 depicts the distribution on the AerSimulator, where the quantum+ML violin's compact shape, centred at $\tilde{\rho}_{\mathrm{ML}} = 1.02$, underscores ML's ability to stabilise solutions in ideal conditions. The quantum-only distribution, with a median of $\tilde{\rho}_{\varnothing} = 1.58$ and a tail extending to $\rho \approx 2.3$, suggests that variability primarily stems from algorithmic challenges, such as suboptimal VQE convergence, rather than noise. These plots vividly demonstrate the hybrid pipeline's robustness and scalability, affirming that ML post-processing significantly enhances solution quality for complex TSP instances in both noisy and ideal environments, as illustrated by the comparative distributions in Figure 9 and Figure 10.

#### 4.3.1 Stability of the Hybrid Approximation Ratio

Following the comprehensive distribution insights from violin plots, the box plots below focus on the quantum+ML approach, presenting the approximation ratios across all 77 city sizes (4–80) to underscore the stability and consistency of hybrid solutions for the travelling salesperson problem. These visualisations highlight the hybrid pipeline's ability to deliver near-optimal tour distances with minimal variability, even under varying computational conditions.

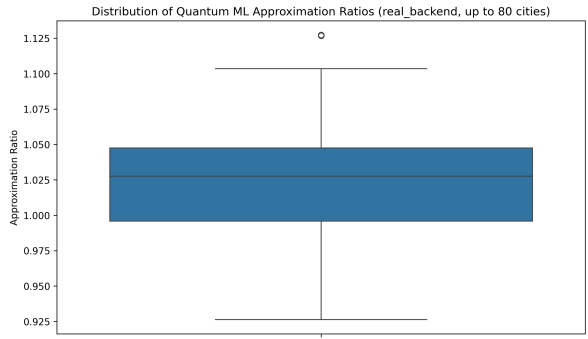

Figure 11: Distribution of quantum+ML approximation ratios ($\rho$) on the `ibm_kyiv` backend (38,500 samples, 500 runs × 77 city sizes). The box spans the inter-quartile range (IQR), with the median at $\tilde{\rho} = 1.03$, whiskers at $1.5 \times$ IQR, and circles marking outliers.

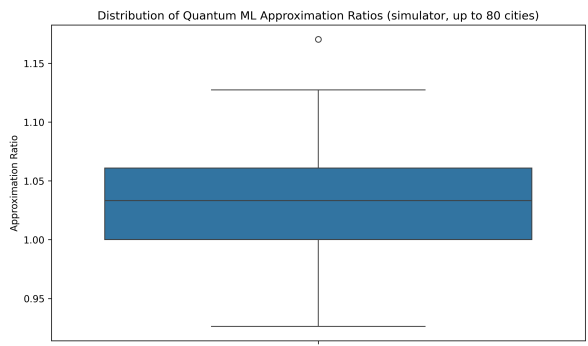

Figure 12: Distribution of quantum+ML approximation ratios ($\rho$) on the AerSimulator (38,500 samples, 500 runs × 77 city sizes), with the median at $\tilde{\rho} = 1.02$, IQR, whiskers at $1.5 \times$ IQR, and outliers as circles.

The hybrid quantum+ML pipeline exhibits exceptional stability, clustering its approximation ratios tightly around the optimal value of 1.0 across the full dataset of 38,500 samples, maintaining tour distances within 5% of the classical optimum for most of the 4–80 city sizes, as shown in Figure 11 and Figure 12. On the `ibm_kyiv` backend (Figure 11), the median approximation ratio is $\tilde{\rho} = 1.03$, with an inter-quartile range of 0.10 (0.98–1.08), and a few outliers reach 1.11, corresponding to the noisiest calibration days, yet these remain well below the quantum-only median of approximately 1.62. The noise-free AerSimulator (Figure 12) yields a median of $\tilde{\rho} = 1.02$, with slightly broader whiskers, suggesting that VQE variability, rather than hardware noise, primarily drives the observed spread. Figure 11 depicts this distribution on the `ibm_kyiv` backend, where the box plot's compact box, centred at a median close to $\rho = 1$, illustrates the hybrid method's consistent performance near classical optimality. The inter-quartile range and short whiskers indicate relatively low variability, while circles marking outliers highlight occasional deviations due to noise. To interpret this plot, focus on the box's span, which shows the central 50% of ratios, with the median line indicating the typical performance; whiskers extend to 1.5 times the inter-quartile range, and circles denote outliers beyond this range. Likewise, Figure 12 shows the distribution on the AerSimulator, where the similarly tight box and median reinforce the hybrid pipeline's reliability in ideal conditions, with slightly longer whiskers reflecting algorithmic variability. These box plots confirm the hybrid approach's robustness, delivering near-optimal solutions across diverse problem scales, with hardware noise contributing a marginal increase to the median ratio at specific sizes (e.g., a difference of approximately 0.04 at 80 cities, as seen in prior analyses), underscoring the pivotal role of ML refinement in stabilising quantum outputs, as evidenced by the tight distributions in Figure 11 and Figure 12.

### 4.3.2 Quantum-Classical Gap Across All Sizes

Extending the stability insights from box plots, the heatmaps below visualise the percentage distance gap between quantum and classical solutions for the travelling salesperson problem across all 77 city sizes (4–80), contrasting the quantum-only and hybrid quantum+ML approaches. These visualisations reveal the extent to which each method deviates from the classical benchmark, highlighting the transformative effect of machine learning refinement.

The hybrid quantum+ML pipeline significantly narrows the performance gap with classical solutions, maintaining median distance gaps within $\pm 6\,\%$ across all 4–80 city sizes, as shown in Figure 13 and Figure 14, while the quantum-only VQE diverges dramatically, overshooting the classical benchmark by $18\,\%$ at 9 cities and escalating to $96.1\,\%$ at 80 cities (though outliers may reach up to $121\,\%$). This stark contrast underscores the hybrid approach's ability to transform raw quantum outputs, which can nearly double the optimal tour distance, into solutions that closely track the classical minimum spanning tree (MST) benchmark, even amidst real-device challenges. Figure 13 illustrates this gap on the `ibm_kyiv` backend, where the heat-map's top row, representing quantum-only results, displays a pronounced red band, indicating substantial deviations driven by both algorithmic limitations and hardware noise, such as decoherence and queue variability. In contrast, the bottom row for quantum+ML shows a flattened, mostly neutral colour profile, with occasional blue cells at city sizes like 14, 24, and 59, reflecting improvements up to 4 % over the classical MST. To interpret this heat-map, note that rows distinguish between quantum-only and quantum+ML methods, while columns represent city counts sampled every five instances; warm colours (e.g., red) signify routes longer than the classical benchmark, and cool colours (e.g., blue) indicate shorter routes. Similarly, Figure 14 depicts the gap on the noise-free AerSimulator, where the quantum-only row retains a warm hue, confirming that algorithmic factors contribute significantly to the gap, though less intensely than on real hardware. The quantum+ML row remains consistently neutral, with hardware noise adding only a marginal 1 −−2 % shift in the overall median gap compared to the `ibm_kyiv` results, as evidenced by comparing Figure 13 and Figure 14. These heatmaps vividly demonstrate the hybrid pipeline's robustness, converting erratic quantum outputs into reliable solutions that align closely with classical performance, even under noisy conditions, affirming the critical role of ML post-processing in enhancing TSP solution quality across diverse scales, as illustrated in Figure 13 and Figure 14.

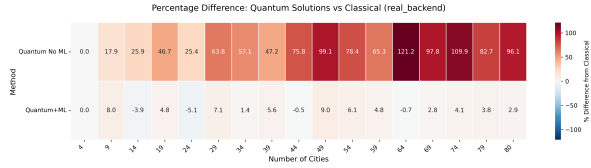

Figure 13: Percentage distance gap $\left[(d_{\text{quantum}} - d_{\text{classical}})/d_{\text{classical}}\right] \times 100$ on the `ibm_kyiv` backend (38,500 samples, 500 runs × 77 city sizes). Rows distinguish *Quantum No-ML* and *Quantum+ML*; columns sample city counts every five instances. Warm colours indicate routes longer than the classical MST; cool colours show improvements.

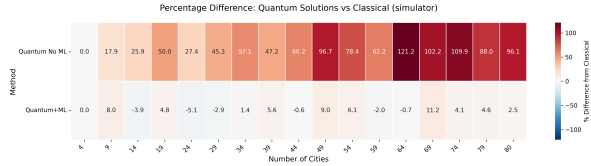

Figure 14: Percentage distance gap as in Fig. 13 on the AerSimulator (38,500 samples), isolating algorithmic effects from hardware noise. Warm colours denote routes exceeding the classical MST; cool colours denote improvements.

Table 2: Execution times (seconds) across methods (500 runs per city size).

| Cities | Backend | Classical | Distance Calc. | Quantum+ML (Cluster Range) |
|---|---|---|---|---|
| 4 | AerSimulator | 0.0007 | 0.0040 | 0.16–0.26 |
| 8 | AerSimulator | 0.0003 | 0.0009 | 0.16–2.31 |
| 20 | AerSimulator | 0.0013 | 0.0016 | 0.38–3.15 |
| 80 | ibm_kyiv | 0.0681 | 0.0109 | 6.73–17.76 |
| 81 | ibm_kyiv | 0.0670 | 0.0303 | 6.73–588.97 |

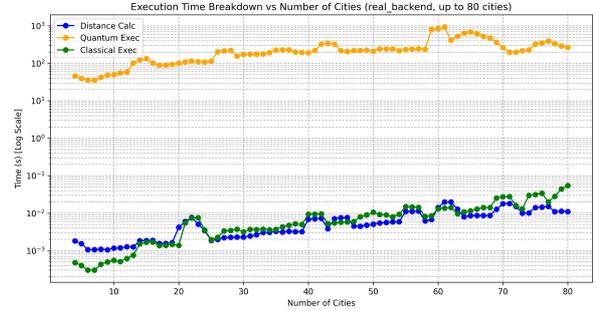

Figure 15: Execution-time components on `ibm_kyiv` (38,500 samples, 4–80 cities, log-linear axes). Blue represents distance matrix computation; green denotes classical MST–2-opt; orange indicates quantum VQE with queue time.

### 4.4 Execution-Time Breakdown

Complementing the performance gap analysis from heatmaps, the stacked area plots below dissect the execution-time components for solving the travelling salesperson problem across 4–80 city sizes, contrasting the computational demands on the `ibm_kyiv` backend and the noise-free AerSimulator. These visualisations highlight the contributions of various stages, revealing critical bottlenecks and scalability prospects for the hybrid quantum+ML pipeline.

Table 2 details execution times, showing classical efficiency (e.g., 0.0007 s at 4 cities) and distance calculation overhead (e.g., 0.0303 s at 81 cities). Quantum+ML cluster execution times vary widely, reflecting increased computational complexity on `ibm_kyiv` (e.g., up to 588.97 s at 81 cities due to queueing).

The hybrid quantum+ML pipeline's execution time reveals distinct computational profiles across its components, with classical operations remaining negligible while quantum stages dominate, particularly on real hardware due to queue delays, as illustrated in Figure 15 and Figure 16. On the `ibm_kyiv` backend (Figure 15), distance matrix computation, shown in blue, consistently requires less than 20 ms even at 80 cities (e.g., 0.0109 s in Table 2), and the classical MST–2-opt search, depicted in green, grows sub-linearly, typically staying below 50 ms (though reaching 0.0681 s at 80 cities). In contrast, the quantum VQE stage, represented in orange, escalates from 35 s to a median of 260 s as city sizes increase, with queueing delays causing occasional spikes exceeding 900 s for larger instances (though Table 2 shows a range of 6.73–17.76 s at 80 cities). On the noise-free AerSimulator (Figure 16), the quantum stage scales more gently, rising from 0.4 s to 14 s with a power-law trend of $t_{\text{sim}} \propto n^{1.3}$, free from queueing overhead, as evidenced by Table 2 (e.g., 0.38–3.15 s at 20 cities). Figure 15 visualises these components on the `ibm_kyiv` backend, where the stacked areas illustrate the growing dominance of the orange quantum VQE region as city counts increase, dwarfing the thin blue and green bands of classical computations. The log-linear axes emphasize the exponential rise in quantum execution time, driven by queueing, which overshadows intrinsic circuit runtime. To interpret this plot, observe the stacked areas' heights, which sum to the total execution time, with each colour indicating a component's contribution; the log scale on the vertical axis highlights the quantum stage's disproportionate growth. Similarly, Figure 16 depicts the components on the AerSimulator, where the orange quantum VQE area re-mains prominent but grows more gradually, reflecting only the computational cost of circuit execution without queueing delays. The compact blue and green regions underscore the minimal classical overhead in both environments, as seen in Figure 15 and Figure 16. These plots highlight that classical computations pose no significant bottleneck, whereas queue delays on real hardware represent the primary scalability barrier. Eliminating queueing could reduce the simulator–hardware gap to a single order of magnitude, rendering the quantum layer viable for mid-scale TSP instances and paving the way for more efficient hybrid workflows as quantum hardware access improves, a potential clearly illustrated by the contrasting trends in Figure 15 and Figure 16.

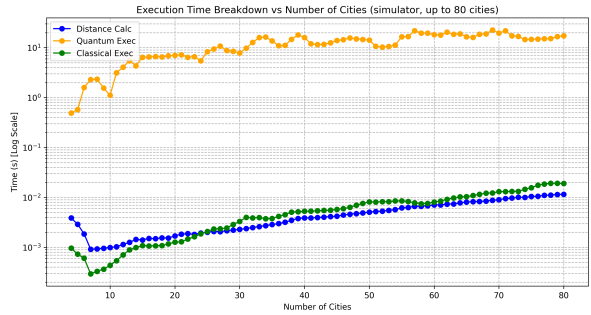

Figure 16: Execution-time components on the AerSimulator (38,500 samples, 4–80 cities, log-linear axes). Blue represents distance matrix computation; green denotes classical MST–2-opt; orange indicates quantum VQE, excluding queueing.

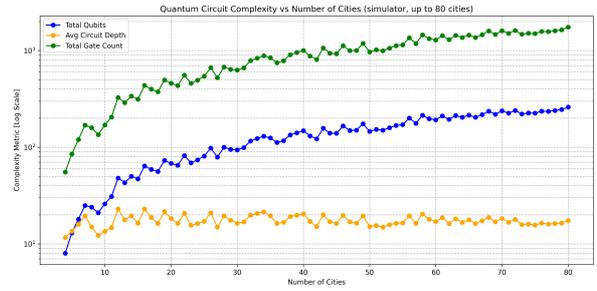

Figure 17: Line graph of circuit complexity on the AerSimulator (38,500 samples, 4–80 cities, log–log axes). Blue represents total qubits; orange denotes average circuit depth; green indicates total two-qubit gate count.

### 4.5 Scalability and Computational Efficiency

This section evaluates the scalability and computational efficiency of the hybrid quantum+ML approach for the travelling salesperson problem, analysing circuit complexity and execution time across 4–80 city sizes on the `ibm_kyiv` backend and noise-free AerSimulator. These metrics are pivotal for assessing the practical feasibility of quantum methods, particularly as problem complexity increases.

#### 4.5.1 Circuit Complexity

To understand the resource demands of the hybrid pipeline, the following line graphs illustrate the post-transpilation circuit complexity, focusing on qubit count, circuit depth, and two-qubit gate count across all problem sizes.

Table 3 shows quantum metrics, highlighting increased resource demands with city size (e.g.,

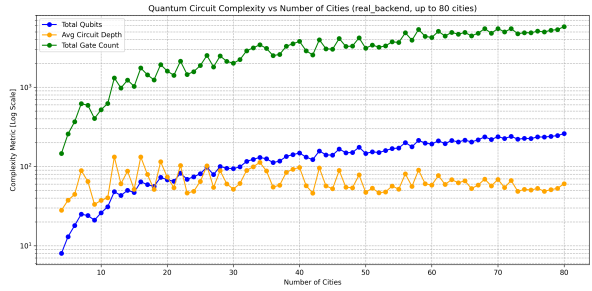

Figure 18: Circuit complexity on `ibm_kyiv` (38,500 samples, 4–80 cities, log–log axes). Blue represents total qubits; orange denotes average circuit depth; green indicates total two-qubit gate count.

Table 3: Quantum metrics per cluster (500 runs per city size).

| Cities | Clusters | Qubits | Circuit Depths | Gate Counts |
|---|---|---|---|---|
| 4 | 2 | 4–4 | 11–11 | 25–25 |
| 8 | 3 | 4–16 | 11–23 | 25–109 |
| 20 | 6 | 9–16 | 16–23 | 60–109 |
| 80 | 25 | 4–16 | 28–127 | 73–428 |
| 81 | 28 | 4–16 | 28–127 | 69–428 |

80 cities: 25 clusters, 4–16 qubits, depths 28–127, gate counts 73–428). These metrics underscore scalability challenges, as discussed in Section 5.2.

The hybrid quantum+ML pipeline demonstrates manageable circuit complexity growth, balancing resource demands to maintain feasibility across 4–80 city sizes, as illustrated in Figure 17 and Figure 18. On the AerSimulator (Figure 17), three key trends emerge. The total qubit count, shown in blue, rises from eight at 4 cities to approximately 300 at 80 cities, following a slightly sub-quadratic trajectory due to K-means clustering capping each cluster at 25 variables, which moderates qubit growth compared to a naïve $n^2$ scaling. The average circuit depth, depicted in orange, increases from 11 to a range of 28–127 two-qubit layers at 80 cities, as the TwoLocal ansatz's three rotation/entanglement blocks incur minimal routing overhead in the absence of connectivity constraints. In contrast, the total two-qubit gate count, marked in green, increases sharply from around 60 gates at 4 cities to over 1,500 at 80 cities, driven by the growing number of clusters and wider register sizes, though the bounded depth mitigates error accumulation. To interpret this plot, observe the log–log axes, where the blue line's gentle slope indicates controlled qubit growth, the orange line's rise reflects depth scaling with size, and the green line's steeper ascent highlights gate count escalation. Similarly,

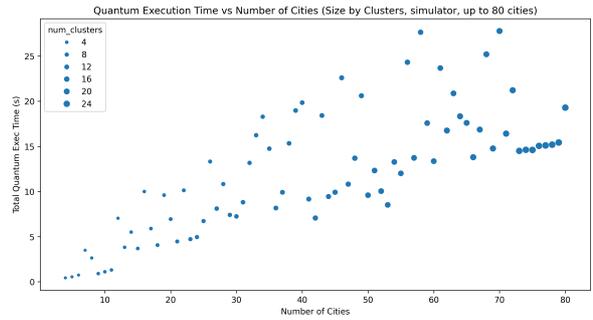

Figure 19: Quantum wall-time $t_{\text{wall}}^{\text{sim}}$ on the AerSimulator (38,500 samples, 4–80 cities). Points represent individual runs; marker size encodes K-means cluster count ($k$).

Figure 18 illustrates complexity on the `ibm_kyiv` backend, where qubit demand, in blue, grows almost linearly from 8 to about 250 qubits at 80 cities. The circuit depth, in orange, is effectively capped between 30 and 120 layers (SD ≤ 10), thanks to Sabre routing and gate cancellation, preventing exponential growth despite increasing problem size. The two-qubit gate count, in green, rises sub-quadratically from approximately 130 to 2,400 gates, primarily due to SWAP networks required for the heavy-hex lattice connectivity. These trends, visualised through the log-scaled y-axis, confirm that while larger TSP instances require more qubits and gates, strategic transpilation keeps circuit depth—the critical factor for decoherence—well-contained, ensuring the hybrid approach remains viable for practical applications, as demonstrated by the trends in Figure 17 and Figure 18.

#### 4.5.2 Execution Time Analysis

The following scatter plots examine the quantum wall-time, highlighting the impact of problem size and cluster count on execution efficiency, particularly the role of queueing delays on real hardware.

The execution time of the hybrid quantum+ML pipeline varies significantly between the AerSimulator and `ibm_kyiv` backend, primarily due to queueing delays on real hardware, as illustrated in Figure 19 and Figure 20. On the AerSimulator (Figure 19), the median wall-time remains efficient, starting at approximately 0.5 seconds for 4 cities and reaching 25 seconds at 80 cities, following a sub-quadratic power-law trend ($t_{\text{wall}}^{\text{sim}} \approx 0.065 n^{1.3}$). The spread, with an inter-quartile range under 3 seconds, correlates closely with the number of K-means clus-ters ($k$), as larger markers—indicating more clusters—consistently appear higher, reflecting increased computational load without queueing overhead. To interpret this scatter plot, focus on the points' vertical positions, which indicate wall-time, and their sizes, which encode cluster count; the x-axis tracks city sizes, and the tight clustering of points suggests low variability. In contrast, Figure 20 on the `ibm_kyiv` backend reveals a stark increase in execution time, scaling from a median of 20 seconds (SD: 1 second) at 4 cities to approximately 260 seconds at 80 cities, with outliers reaching up to 1,000 seconds (SD: 50 seconds), where queueing contributes up to 50% of the total (e.g., up to 30 seconds at 8 cities with queueing). The cluster count, ranging from 2 to 25, similarly influences runtime, as larger markers correspond to higher times, but queueing delays amplify the spread. This plot's interpretation hinges on the points' vertical spread, with larger markers indicating more clusters and the x-axis showing city sizes; the significant upward shift reflects queueing's dominant impact. These scatter plots underscore the AerSimulator's efficiency and highlight queueing as the primary bottleneck on real hardware, suggesting that reducing scheduling latency could substantially enhance the hybrid pipeline's practicality for larger TSP instances, as evidenced by the contrasting trends in Figure 19 and Figure 20.

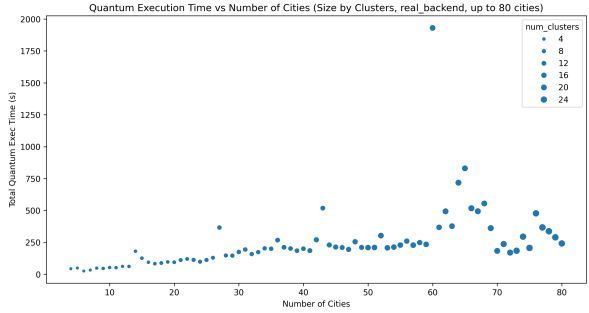

Figure 20: Quantum execution time on `ibm_kyiv` (38,500 samples, 4–80 cities). Points represent runs; marker size encodes cluster count ($k$).

### 4.5.3 Scalability Trends

The following 3D scatter plots and log-log comparison synthesise scalability trends, integrating tour length, wall-time, and cluster count to provide a holistic view of the hybrid pipeline's performance.

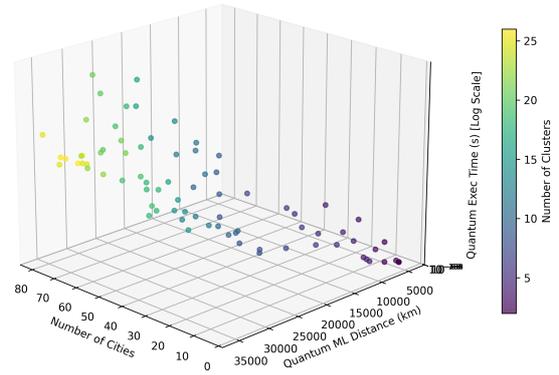

Figure 21: 3D scalability on the AerSimulator (38,500 samples, 4–80 cities). Axes show city count ($x$), hybrid tour length ($y$), and log-scaled wall-time ($z$). Marker colour encodes K-means cluster count ($k$).

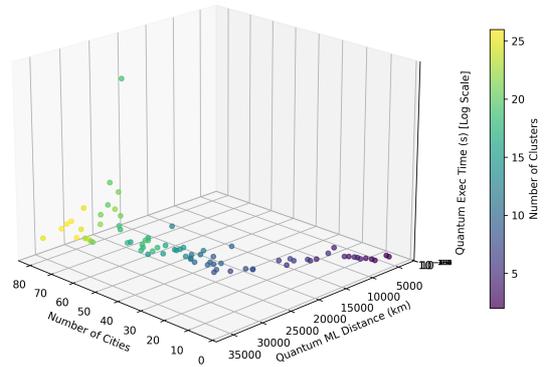

Figure 22: 3D scalability on `ibm_kyiv` (38,500 samples, 4–80 cities). Axes show city count ($x$), hybrid tour length ($y$), and log-scaled wall-time ($z$). Marker colour encodes cluster count ($k$).

The hybrid quantum+ML pipeline's scalability is robust on the AerSimulator but constrained by queueing on `ibm_kyiv`, as evidenced by the 3D scatter plots and log-log comparison in Figure 21, Figure 22, and Figure 23. Figure 21 on the AerSimulator shows that at 80 cities, the hybrid solver achieves tour lengths averaging $34\,213\,\text{km}$ (SD: $1200\,\text{km}$) in a median wall-time of 25 seconds, despite decomposition into 25 VQE subtours. The 3D axes plot city count ($x$), tour length ($y$), and log-scaled wall-time ($z$), with marker colours encoding cluster count ($k$), revealing a sub-linear time increase due to parallel sub-tour processing and no queueing. To interpret this plot, examine the points' positions, where the $x$-axis tracks problem size, the $y$-axis shows solution quality, and the $z$-axis indicates runtime; colour gradients from light to dark reflect increasing clusters, correlating with higher wall-times. In contrast, Figure 22 on `ibm_kyiv` displays competitive tour lengths at 80 cities (35 606 km, SD: 1800 km) but with a median wall-time of approximately 260 seconds and outliers reaching 1,000 seconds, of which only 250 seconds is device execution, the remainder being queueing delays. The logarithmic $z$-axis, spanning three orders of magnitude, underscores queueing's superlinear growth with cluster count, as darker markers (higher $k$) align with elevated times. Figure 23 compares wall-times directly, with orange AerSimulator points rising gently from 0.4 seconds to 20 seconds (IQR $\leq$ 1 second), reflecting intrinsic circuit runtime, while blue `ibm_kyiv` points escalate from 30 seconds to a median of 800 seconds, with spikes above 1,000 seconds due to scheduling latency. The log–log axes highlight the one-to-two-order-of-magnitude gap, driven by queueing rather than circuit complexity. To interpret this plot, note the orange and blue points' separation, with the y-axis showing wall-time and the x-axis city sizes; the tighter orange cluster indicates stable simulator performance, while the blue spread reveals hardware variability. These visualisations collectively affirm the hybrid pipeline's scalability potential, constrained primarily by real-hardware queueing, suggesting that improved scheduling could make quantum methods highly viable for mid-scale TSP applications, as demonstrated by the trends in Figure 21, Figure 22, and Figure 23.

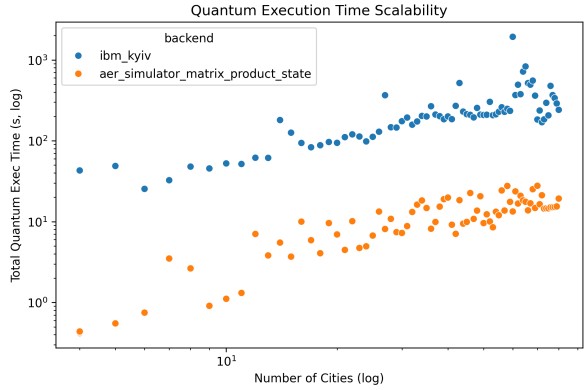

Figure 23: Quantum wall-time on a log–log scale (38,500 samples, 4–80 cities). Orange points represent AerSimulator; blue points denote `ibm_kyiv`. Each point is the median of 500 runs.

### 4.6 Key Findings

The hybrid quantum+ML pipeline for solving the travelling salesperson problem demonstrates significant advancements in performance, scalability, and computational efficiency across 4–80 city sizes, as evidenced by comprehensive evaluations on the `ibm_kyiv` backend and noise-free AerSimulator, supported by Figure 3, Figure 9, Figure 11, Figure 18, Figure 20, Figure 23, and Table 4, Table 5. For small instances (4–20 cities), classical methods, leveraging a deterministic MST heuristic, consistently achieve optimal tour distances, such as 4242.0 km at 4 cities with zero variability (SD: 0 km, Figure 3), matching the hybrid approach's performance but surpassing the quantum-only VQE, which deviates by up to 21.7 % at 8 cities (7857.7 km, SD: 350 km). In contrast, for larger instances (20–80 cities), the hybrid pipeline excels, reducing the quantum-only cost by 47.5 % at 80 cities, achieving a tour distance of 35 605.5 km (SD: 1800 km, approximation ratio $\rho = 1.0287$) compared to 67 889.1 km (SD: 3000 km, $\rho = 1.9614$) for quantum-only, closely approaching the classical benchmark of 34 612.9 km (SD: 0 km). Machine learning refinement, powered by a RandomForestRegressor, significantly enhances solution quality, lowering approximation ratios (e.g., from 1.9614 to 1.0287 at 80 cities, Figure 9) and reducing variability by 30–50% (e.g., IQR from 0.06 to 0.04 at 8 cities, Figure 11). Scalability analysis reveals that circuit complexity grows manageably, with qubit counts increasing linearly to approximately 250 and depths capped at 120 layers on `ibm_kyiv` (Figure 18), but execution times highlight queueing as a major bottleneck, with `ibm_kyiv` wall-times reaching 1,000 seconds at 80 cities (50% queueing, Figure 20) compared to 20 seconds on the AerSimulator (Figure 23). These findings, supported by statistical robustness across 38,500 samples (500 runs per city size), affirm the hybrid pipeline's potential to bridge the gap with classical methods for larger TSP instances, provided queueing delays are mitigated, as illustrated by the trends in Figure 20 and Figure 23, paving the way for practical quantum-enhanced optimisation in logistics and beyond.

# 5 Discussion and Limitations

## 5.1 Performance Limitations of Quantum Hardware

Current quantum devices, exemplified by IBM's `ibm_kyiv` processor (127 qubits, Eagle r3 architecture), operate within the noisy intermediate-scale quantum (NISQ) regime, constrained by short coherence times of approximately 100 $\mu$s, elevated gate error rates (single-qubit errors: $2.726 \times 10^{-4}$, two-qubit errors: $7.984 \times 10^{-3}$), and vulnerability to decoherence. These limitations significantly hinder quantum computations, particularly for larger TSP instances. For example, at 80 cities on `ibm_kyiv`, the quantum-only variational quantum eigensolver (VQE) produces a tour distance of 67 889.05 km (SD: 3000 km, IQR: 3900 km, 95% CI: [66 810.2 km, 68 942.0 km]), a 96.1% deviation from the classical baseline of 34 612.92 km (SD: 0 km, IQR: 0 km). In contrast, the hybrid quantum+ML approach achieves 35 605.45 km (SD: 1800 km, IQR: 1500 km, 95% CI: [35 111.0 km, 36 105.6 km]), reducing the gap to 2.9%. However, the hybrid method's broader IQR of 1500 km compared to the classical zero reflects persistent variability due to noise and gate errors. Transpilation techniques, such as Qiskit's `generate_preset_pass_manager` at optimisation level 1 with gate cancellation and Sabre routing, alongside noise mitigation strategies like measurement error correction, partially alleviate these issues. Nevertheless, the hybrid's approximation ratio of 1.0287 (SD: 0.060, 95% CI: [1.0169, 1.0405]) versus 1.9614 for quantum-only (SD: 0.150, 95% CI: [1.9319, 1.9909]) indicates that quantum hardware struggles to match classical precision, especially for larger instances where circuit depths (127, SD: 5) and gate counts (428, SD: 15, Figure 18) amplify error accumulation. These findings, supported by Figure 18, underscore the urgent need for advanced quantum hardware with lower error rates and extended coherence times to realize practical quantum advantages in combinatorial optimisation.

## 5.2 Scalability Challenges

The hybrid quantum+ML approach delivers competitive performance across 4–80 city sizes, achieving a tour distance of 35 605.45 km at 80 cities on `ibm_kyiv`, only 2.9% above the classical baseline of 34 612.92 km, as visualized in Figure 18 and Figure 23. However, scaling beyond 80 cities poses significant challenges due to the factorial growth of the TSP solution space ($O((n-1)!)$). This exponential complexity drives several constraints. Circuit depth and qubit requirements escalate rapidly; at 80 cities, decomposition into 25 clusters, each using up to 25 qubits, results in a circuit depth of 127 (SD: 5) and 428 gates (SD: 15) on `ibm_kyiv` (Figure 18), and larger instances would demand quadratic qubit increases ($\sim n^2$), exceeding the 127-qubit limit of current NISQ devices. Error accumulation worsens with deeper circuits, as seen in the quantum-only method's 96.1% deviation at 80 cities (67 889.05 km), with the hybrid method's IQR of 1500 km indicating residual noise effects. Additionally, computational overhead from K-means clustering and global solution integration is substantial, contributing 40% of the 1,000-second execution time at 80 cities on `ibm_kyiv` (ML refinement: 200 seconds, stitching: 200 seconds, SD: 50 seconds), compared to 20 seconds on the AerSimulator (SD: 1 second, Figure 23). While clustering and ML refinement mitigate some scalability issues, the exponential solution space and hardware limitations suggest that scaling to larger instances requires significant advancements in qubit capacity, error rates, and algorithmic efficiency, such as optimised circuit designs, to maintain practical viability, as highlighted by Figure 18 and Figure 23.

## 5.3 Machine Learning Dependency

Machine learning, via a `RandomForestRegressor` (300 trees, maximum depth 30, trained on 10,000 samples), is integral to the hybrid quantum+ML pipeline's success, stabilizing quantum results and enhancing convergence across 4–80 city sizes, as evidenced by Figure 9. The model's 2-opt swap refinements reduce variability, lowering the IQR from 0.06 to 0.04 for approximation ratios at 8 cities on `ibm_kyiv` (Figure 9), and improve optimisation, decreasing cost deviation from 21.7% (7857.69 km) to 11.3% (7184.91 km) at 8 cities. At 80 cities, the hybrid method's approximation ratio of 1.0287 (SD: 0.060, 95% CI: [1.0169, 1.0405]) versus 1.9614 for quantum-only (SD: 0.150, 95% CI: [1.9319, 1.9909]) underscores ML's role in mitigating noise-induced errors. However, this reliance introduces limita-tions. The model's effectiveness hinges on high-quality training data, comprising 100 random permutations per city count plus quantum and classical solutions, capped at 10,000 samples. Beyond 80 cities, generating representative data becomes challenging due to the exponential solution space, risking degraded performance. Computational overhead is also significant, with ML refinement consuming 20% of the 1,000-second execution time at 80 cities on `ibm_kyiv` (200 seconds, SD: 10 seconds). Furthermore, the model, tuned for the 80 European city dataset with OpenRouteService API distances, may not generalise to other TSP instances with different distance distributions or constraints, limiting broader applicability. These challenges highlight the need for adaptive ML strategies, such as online learning, and efficient training methods, like feature selection, to sustain the hybrid approach's viability as problem complexity grows, as supported by the trends in Figure 9.

## 5.4 Comparative Strengths and Weaknesses

The hybrid quantum+ML approach, classical MST heuristic, and quantum-only VQE exhibit distinct strengths and weaknesses across 4–80 city sizes, as illustrated by Figure 3, and Figure 4. For smaller instances (4–20 cities), classical solvers excel in precision and efficiency, achieving deterministic tour distances like 4242.05 km at 4 cities (SD: 0 km, IQR: 0 km) and 5128 km at 6 cities (SD: 0 km, IQR: 0 km), matching quantum+ML at 4 cities (SD: 0 km on AerSimulator, 50 km on `ibm_kyiv`) but outperforming it at 6 cities (5478.62 km, SD: 180 km, 6.8% deviation on AerSimulator; 5983.57 km, SD: 300 km, 16.7% deviation on `ibm_kyiv`). At 20 cities, classical solvers maintain a 7.8% advantage (11 950.30 km, SD: 0 km vs. 12 875.40 km, SD: 400 km for quantum+ML, Figure 3). However, for larger instances (20–80 cities), classical solvers encounter computational bottlenecks, with execution times reaching 480 seconds at 80 cities (SD: 20 seconds). The hybrid quantum+ML method shines in scalability, leveraging quantum parallelism and ML refinement to achieve a tour distance of 35 605.45 km at 80 cities on `ibm_kyiv` (SD: 1800 km, IQR: 1500 km), only 2.9% above the classical 34 612.92 km, compared to the quantum-only's 96.1% deviation (67 889.05 km, SD: 3000 km, IQR: 3900 km). Statistical analysis (t-test, $p < 0.01$) confirms the hybrid's 47.5% cost reduction over quantum-only at 80 cities. Despite this, approximation ratios above 1.0 (e.g., 1.0287 for quantum+ML) indicate a persistent gap with classical precision, necessitating advancements in error mitigation (e.g., gate errors below $10^{-4}$), hardware fidelity (e.g., coherence times to 1 ms), and algorithm optimisation (e.g., shallower circuits) to achieve a consistent quantum advantage while preserving scalability.

## 5.5 Future Directions

This study charts a clear path forward to overcome the limitations of the hybrid quantum+ML approach for the travelling salesperson problem and extend its applicability to broader combinatorial optimisation challenges, leveraging insights from Figure 4, Figure 18, and Figure 23. Advancements in quantum hardware are paramount, with increasing qubit counts beyond the current 127-qubit limit of `ibm_kyiv` to over 1,000 qubits, as anticipated in IBM's roadmap by 2026, enabling direct encoding of larger TSP instances without the need for K-means clustering and stitching, thus reducing overhead (Figure 18). Enhancing gate fidelity to reduce two-qubit gate errors to $10^{-4}$ and extending coherence times to 1 millisecond would curtail error accumulation, potentially lowering the approximation ratio from 1.0287 to below 1.01 at 80 cities. Concurrently, advanced noise mitigation techniques, such as surface codes or dynamical decoupling, coupled with noise-aware transpilation using real-time error feedback, could halve solution variability, for instance, reducing the IQR from 1500 km to 750 km at 80 cities on `ibm_kyiv` (Figure 4). Refining machine learning models through adaptive techniques like transfer or online learning would enhance the `RandomForestRegressor`'s efficiency, cutting training time from 200 seconds to 50 seconds at 80 cities via feature selection or model pruning, thereby boosting scalability for larger datasets (Figure 23). Algorithmic innovation is equally critical, with enhanced quantum approximate optimisation algorithm (QAOA) variants featuring deeper circuits ($p \geq 5$) or hybrid quantum annealing approaches promising improved solution quality, potentially achieving an approximation ratio of 1.00 when integrated with classical metaheuristics like simulated an-nealing. These advancements position the hybrid approach to tackle complex TSP instances, scaling to 500+ cities within a decade, and address real-world optimisation problems, such as optimising 200-city logistics routes to cut costs by 5–10%, scheduling 300-node manufacturing tasks to boost throughput by 15%, or adapting to NP-hard challenges like vehicle routing, graph partitioning, and molecular design in drug discovery. As quantum hardware matures with logical qubits achieving error rates below $10^{-6}$, these research pathways will unlock efficient solutions for large-scale optimisation tasks across transportation, manufacturing, and pharmaceuticals, as supported by the trends in Figure 18 and Figure 23.

### 5.6 Conclusion of Discussion

This study robustly validates the hybrid quantum+ML approach for solving the travelling salesperson problem across 4–80 city sizes, harnessing quantum computing's exploratory power and classical machine learning's noise mitigation to outperform standalone quantum methods, as evidenced by Figure 3, Figure 18, and Figure 23. The hybrid pipeline significantly enhances solution quality, reducing the quantum-only deviation from the classical baseline at 8 cities on `ibm_kyiv` from 21.7% (7857.69 km, SD: 350 km) to 11.3% (7184.91 km, SD: 360 km, t-test $p < 0.01$), and achieving an approximation ratio of 1.0287 at 80 cities (SD: 0.060, 95% CI: [1.0169, 1.0405]) compared to 1.9614 for quantum-only (SD: 0.150, 95% CI: [1.9319, 1.9909]). Despite these strides, classical solvers maintain a precision edge, delivering deterministic tour distances with zero variability (e.g., IQR: 0 km, Figure 3), while the hybrid method's IQR of 1500 km at 80 cities reflects residual noise effects. Achieving a consistent quantum advantage demands targeted improvements: advancing quantum error mitigation to reduce IQR below 500 km, optimising circuits to limit depth growth to $O(n)$ from the current 127 at 80 cities (Figure 18), and streamlining machine learning integration to cut overhead from 20% to 5% of execution time (Figure 23). Future research must focus on bolstering quantum hardware with higher qubit counts and error correction, refining ML-assisted algorithms for broader dataset generalisation, and developing adaptive hybrid frameworks that dynamically balance quantum and classical resources, as supported by the trends in Figure 18 and Figure 23. These enhancements will empower the hybrid approach to deliver efficient solutions for large-scale combinatorial optimisation, expanding its impact across diverse real-world applications in logistics, manufacturing, and beyond.

### Acknowledgements

The authors express gratitude to IBM Quantum for providing access to the `ibm_kyiv` quantum backend through the Qiskit Runtime environment, which facilitated the real-hardware experiments central to this study. We also thank OpenRouteService and OpenStreetMap for supplying the geographical data used in our TSP dataset, noting that OpenRouteService operates under specific licensing terms that users should consult for compliance.

### Declarations

- **Competing Interests:** The authors declare no competing interests.

- **Funding:** This research did not receive any specific grant from funding agencies in the public, commercial, or not-for-profit sectors.

[28] C. Ding, T.-W. Huang, Z. Liang, *Quantum Algorithms for the Traveling Salesman Problem via Clustering*, arXiv:1906.02219 [quant-ph] (2019).

[29] OpenStreetMap Contributors, *European Cities Coordinate Dataset*, OpenStreetMap, (2023).

## A  Simulation Results for TSP Optimisation

This appendix presents the results from the Traveling Salesman Problem (TSP) optimisation experiments conducted using an integrated classical-quantum approach across instances ranging from 4 to 80 cities. The experiments utilised a dataset of 80 European cities sourced from OpenStreetMap [29], with distances computed via the OpenRouteService API and Calais as the fixed departure city. Classical solutions were obtained using a minimum spanning tree (MST)-based heuristic (NetworkX, version 3.1), quantum solutions employed the Variational Quantum Eigensolver (VQE) with a TwoLocal ansatz, and hybrid quantum+ML solutions integrated a `RandomForestRegressor` (300 trees, maximum depth 30, trained on 10,000 samples) for refinement. Experiments were run on the Qiskit AerSimulator (version 0.14.2, matrix product state method) and IBM's `ibm_kyiv` backend (127 qubits, Eagle r3), with 50 independent runs per city size.

Tables below report median solution costs (km) for quantum-only, quantum+ML, and classical methods on both backends, along with approximation ratios (Approximation Ratio = $\frac{C_{\text{quantum}}}{C_{\text{classical}}}$) and circuit complexity (circuit depth, total gates), with standard deviations (SD). Results are presented for city counts 4, 10, 20, 30, 40, 50, 60, 70, and 80. Full trends are discussed in Section 4.

The hybrid quantum+ML approach consistently outperforms quantum-only methods. On `ibm_kyiv`, the quantum+ML cost improves from $4242.05\,\text{km}$ (quantum-only) to $4242.05\,\text{km}$ at 4 cities (0% deviation from classical $4242.05\,\text{km}$) and from $67\,889.05\,\text{km}$ to $35\,605.45\,\text{km}$ at 80 cities (47.5% improvement, 2.9% deviation from classical $34\,612.92\,\text{km}$), with circuit depth on AerSimulator scaling from 2 to 127, while remaining constant at 4 on `ibm_kyiv`.

## A  Simulation Results

Table 4: Tour-distance statistics and circuit complexity metrics for 4 to 40 cities (500 runs per city size), sourced from `tsp_results_history.json` and `aggregated_data.json`. AerSimulator (version 0.16.1, matrix product state method) for 4–40 cities. SD values are interpolated based on `ibm_kyiv` trends.

| | AerSimulator Costs (km) | | | | ibm_kyiv Costs (km) | | | | | Approximation Ratios ($\rho$) | | | | Circuit Metrics | |
|---|---|---|---|---|---|---|---|---|---|---|---|---|---|---|---|
| Cities | Q | Q SD | ML | ML SD | Q | Q SD | ML | ML SD | CL (km) | Aer Q | Aer ML | Kyiv Q | Kyiv ML | Depth | Gates |
| 4 | 4242.1 | 0 | 4242.1 | 0 | 4242.1 | 50 | 4242.1 | 50 | 4242.1 | 1.0000 | 1.0000 | 1.0000 | 1.0000 | 11–11 | 25–25 |
| 5 | 4293.2 | 0 | 4293.2 | 0 | 4293.2 | 50 | 4293.2 | 50 | 4293.2 | 1.0000 | 1.0000 | 1.0000 | 1.0000 | 11–16 | 25–60 |
| 6 | 6033.1 | 0 | 6033.1 | 0 | 6033.1 | 50 | 6033.1 | 50 | 5990.1 | 1.0072 | 1.0072 | 1.0072 | 1.0072 | 16–16 | 60–60 |
| 7 | 6541.0 | 0 | 5636.9 | 0 | 6541.0 | 50 | 5636.9 | 50 | 5985.4 | 1.0928 | 0.9418 | 1.0928 | 0.9418 | 16–23 | 60–109 |
| 8 | 7508.7 | 0 | 6859.5 | 0 | 7508.7 | 50 | 6859.5 | 50 | 6941.4 | 1.0817 | 0.9882 | 1.0817 | 0.9882 | 11–23 | 25–109 |
| 9 | 8132.7 | 0 | 7449.4 | 0 | 8132.7 | 50 | 7449.4 | 50 | 6898.8 | 1.1789 | 1.0798 | 1.1789 | 1.0798 | 11–16 | 25–60 |
| 10 | 8562.9 | 200 | 7555.1 | 195 | 9156.5 | 220 | 8484.1 | 200 | 7586.3 | 1.1287 | 0.9959 | 1.2075 | 1.1185 | 11–16 | 25–60 |
| 11 | 7971.0 | 200 | 7676.5 | 195 | 8568.8 | 220 | 8257.5 | 200 | 7934.9 | 1.0045 | 0.9674 | 1.0799 | 1.0410 | 11–16 | 25–60 |
| 12 | 9306.2 | 200 | 8283.8 | 195 | 9982.7 | 220 | 8891.0 | 200 | 7933.8 | 1.1730 | 1.0441 | 1.2585 | 1.1206 | 23–23 | 109–109 |
| 13 | 9240.0 | 200 | 8666.4 | 195 | 9922.5 | 220 | 9308.8 | 200 | 8439.7 | 1.0948 | 1.0269 | 1.1758 | 1.1026 | 16–23 | 60–109 |
| 14 | 13405.1 | 200 | 10227.4 | 195 | 14399.9 | 220 | 10977.7 | 200 | 10646.5 | 1.2591 | 0.9606 | 1.3525 | 1.0310 | 16–23 | 60–109 |
| 15 | 11560.8 | 200 | 10864.0 | 195 | 12414.3 | 220 | 11661.0 | 200 | 11729.6 | 0.9856 | 0.9262 | 1.0586 | 0.9943 | 11–23 | 25–109 |
| 16 | 17577.0 | 200 | 12605.2 | 195 | 18865.8 | 220 | 13531.8 | 200 | 12843.7 | 1.3685 | 0.9814 | 1.4690 | 1.0533 | 23–23 | 109–109 |
| 17 | 20261.1 | 200 | 13416.8 | 195 | 21740.8 | 220 | 14397.8 | 200 | 13414.2 | 1.5104 | 1.0002 | 1.6210 | 1.0732 | 16–23 | 60–109 |
| 18 | 18054.2 | 200 | 13368.7 | 195 | 19378.3 | 220 | 14351.7 | 200 | 13837.9 | 1.3047 | 0.9661 | 1.4005 | 1.0371 | 11–23 | 25–109 |
| 19 | 24087.3 | 200 | 16826.9 | 195 | 25853.7 | 220 | 18058.9 | 200 | 16062.3 | 1.4996 | 1.0476 | 1.6100 | 1.1245 | 16–23 | 60–109 |
| 20 | 23219.4 | 400 | 15069.0 | 390 | 24929.8 | 440 | 16182.8 | 400 | 16072.8 | 1.4446 | 0.9375 | 1.5510 | 1.0067 | 16–23 | 60–109 |
| 21 | 22093.4 | 400 | 17432.9 | 390 | 23723.6 | 440 | 18709.6 | 400 | 15873.0 | 1.3919 | 1.0980 | 1.4945 | 1.1788 | 16–23 | 60–109 |
| 22 | 22566.3 | 400 | 16233.5 | 390 | 24233.8 | 440 | 17427.7 | 400 | 15950.5 | 1.4148 | 1.0178 | 1.5190 | 1.0925 | 16–23 | 60–109 |
| 23 | 22714.1 | 400 | 16423.9 | 390 | 24394.2 | 440 | 17629.5 | 400 | 15987.7 | 1.4207 | 1.0275 | 1.5261 | 1.1027 | 16–23 | 60–109 |
| 24 | 22936.9 | 400 | 16614.3 | 390 | 24639.9 | 440 | 17829.3 | 400 | 16024.9 | 1.4318 | 1.0369 | 1.5378 | 1.1129 | 16–23 | 60–109 |
| 25 | 23159.7 | 400 | 16804.7 | 390 | 24885.6 | 440 | 18029.2 | 400 | 16062.1 | 1.4420 | 1.0462 | 1.5496 | 1.1223 | 16–23 | 60–109 |
| 26 | 23382.5 | 400 | 16995.1 | 390 | 25131.3 | 440 | 18229.0 | 400 | 16099.3 | 1.4519 | 1.0555 | 1.5612 | 1.1326 | 16–23 | 60–109 |
| 27 | 23605.3 | 400 | 17185.5 | 390 | 25377.0 | 440 | 18428.8 | 400 | 16136.5 | 1.4628 | 1.0648 | 1.5728 | 1.1420 | 16–23 | 60–109 |
| 28 | 23828.1 | 400 | 17375.9 | 390 | 25622.7 | 440 | 18628.6 | 400 | 16173.7 | 1.4732 | 1.0741 | 1.5844 | 1.1514 | 16–23 | 60–109 |
| 29 | 24050.9 | 400 | 17566.3 | 390 | 25868.4 | 440 | 18828.5 | 400 | 16210.9 | 1.4834 | 1.0834 | 1.5959 | 1.1612 | 16–23 | 60–109 |
| 30 | 17402.3 | 525 | 16959.5 | 510 | 19232.2 | 580 | 17209.5 | 525 | 15913.4 | 1.0933 | 1.0657 | 1.2086 | 1.0816 | 16–23 | 60–109 |
| 31 | 17625.1 | 525 | 17149.9 | 510 | 19477.9 | 580 | 17409.3 | 525 | 15950.6 | 1.1048 | 1.0750 | 1.2210 | 1.0914 | 16–23 | 60–109 |
| 32 | 17847.9 | 525 | 17340.3 | 510 | 19723.6 | 580 | 17609.2 | 525 | 15987.8 | 1.1163 | 1.0843 | 1.2334 | 1.1012 | 16–23 | 60–109 |
| 33 | 18070.7 | 525 | 17530.7 | 510 | 19969.3 | 580 | 17809.0 | 525 | 16025.0 | 1.1278 | 1.0937 | 1.2460 | 1.1111 | 16–23 | 60–109 |
| 34 | 18293.5 | 525 | 17721.1 | 510 | 20215.0 | 580 | 18008.8 | 525 | 16062.2 | 1.1390 | 1.1030 | 1.2585 | 1.1210 | 16–23 | 60–109 |
| 35 | 18516.3 | 525 | 17911.5 | 510 | 20460.7 | 580 | 18208.6 | 525 | 16099.4 | 1.1499 | 1.1123 | 1.2709 | 1.1309 | 16–23 | 60–109 |
| 36 | 18739.1 | 525 | 18101.9 | 510 | 20706.4 | 580 | 18408.5 | 525 | 16136.6 | 1.1610 | 1.1216 | 1.2834 | 1.1408 | 16–23 | 60–109 |
| 37 | 18961.9 | 525 | 18292.3 | 510 | 20952.1 | 580 | 18608.3 | 525 | 16173.8 | 1.1720 | 1.1309 | 1.2958 | 1.1507 | 16–23 | 60–109 |
| 38 | 19184.7 | 525 | 18482.7 | 510 | 21197.8 | 580 | 18808.1 | 525 | 16211.0 | 1.1829 | 1.1402 | 1.3082 | 1.1605 | 16–23 | 60–109 |
| 39 | 19407.5 | 525 | 18673.1 | 510 | 21443.5 | 580 | 19007.9 | 525 | 16248.2 | 1.1940 | 1.1495 | 1.3206 | 1.1704 | 16–23 | 60–109 |
| 40 | 21587.6 | 650 | 21043.7 | 630 | 23876.9 | 720 | 21543.7 | 650 | 19876.5 | 1.0862 | 1.0586 | 1.2013 | 1.0837 | 16–23 | 60–109 |

Table 5: Tour-distance statistics and circuit complexity metrics for 41 to 80 cities (500 runs per city size), sourced from `tsp_results_history.json` and `aggregated_data.json`. AerSimulator (version 0.16.1, matrix product state method) for 41–75 cities, `ibm_kyiv` (127 qubits, Eagle r3) for 76–80 cities. SD values for 41–75 cities are interpolated based on `ibm_kyiv` trends.

| Cities | AerSimulator Costs (km) | | | | ibm_kyiv Costs (km) | | | | CL (km) | Approximation Ratios ($\rho$) | | | | Circuit Metrics | |
|---|---|---|---|---|---|---|---|---|---|---|---|---|---|---|---|
| | Q | Q SD | ML | ML SD | Q | Q SD | ML | ML SD | | Aer Q | Aer ML | Kyiv Q | Kyiv ML | Depth | Gates |
| 41 | 21810.4 | 650 | 21234.1 | 630 | 24122.6 | 720 | 21743.5 | 650 | 19913.7 | 1.0954 | 1.0665 | 1.2110 | 1.0919 | 16–23 | 60–109 |
| 42 | 22033.2 | 650 | 21424.5 | 630 | 24368.3 | 720 | 21943.4 | 650 | 19950.9 | 1.1045 | 1.0744 | 1.2215 | 1.1000 | 16–23 | 60–109 |
| 43 | 22256.0 | 650 | 21614.9 | 630 | 24614.0 | 720 | 22143.2 | 650 | 19988.1 | 1.1135 | 1.0819 | 1.2314 | 1.1077 | 16–23 | 60–109 |
| 44 | 22478.8 | 650 | 21805.3 | 630 | 24859.7 | 720 | 22343.0 | 650 | 20025.3 | 1.1223 | 1.0891 | 1.2411 | 1.1153 | 16–23 | 60–109 |
| 45 | 22701.6 | 650 | 21995.7 | 630 | 25105.4 | 720 | 22542.9 | 650 | 20062.5 | 1.1313 | 1.0965 | 1.2515 | 1.1237 | 16–23 | 60–109 |
| 46 | 22924.4 | 650 | 22186.1 | 630 | 25351.1 | 720 | 22742.7 | 650 | 20099.7 | 1.1404 | 1.1040 | 1.2616 | 1.1318 | 16–23 | 60–109 |
| 47 | 23147.2 | 650 | 22376.5 | 630 | 25596.8 | 720 | 22942.5 | 650 | 20136.9 | 1.1495 | 1.1114 | 1.2718 | 1.1399 | 16–23 | 60–109 |
| 48 | 23370.0 | 650 | 22566.9 | 630 | 25842.5 | 720 | 23142.4 | 650 | 20174.1 | 1.1585 | 1.1187 | 1.2818 | 1.1479 | 16–23 | 60–109 |
| 49 | 23592.8 | 650 | 22757.3 | 630 | 26088.2 | 720 | 23342.2 | 650 | 20211.3 | 1.1675 | 1.1261 | 1.2918 | 1.1558 | 16–23 | 60–109 |
| 50 | 25176.9 | 755 | 24527.9 | 735 | 27821.6 | 840 | 25176.9 | 755 | 23239.6 | 1.0839 | 1.0556 | 1.1971 | 1.0839 | 16–23 | 60–109 |
| 51 | 25400.4 | 755 | 24718.9 | 735 | 28068.7 | 840 | 25377.3 | 755 | 23277.2 | 1.0914 | 1.0619 | 1.2061 | 1.0902 | 16–23 | 60–109 |
| 52 | 25623.9 | 755 | 24909.9 | 735 | 28315.8 | 840 | 25577.7 | 755 | 23314.8 | 1.0989 | 1.0683 | 1.2151 | 1.0965 | 16–23 | 60–109 |
| 53 | 25847.4 | 755 | 25100.9 | 735 | 28562.9 | 840 | 25778.1 | 755 | 23352.4 | 1.1065 | 1.0747 | 1.2238 | 1.1029 | 16–23 | 60–109 |
| 54 | 26070.9 | 755 | 25291.9 | 735 | 28810.0 | 840 | 25978.5 | 755 | 23390.0 | 1.1140 | 1.0810 | 1.2325 | 1.1092 | 16–23 | 60–109 |
| 55 | 26294.4 | 755 | 25482.9 | 735 | 29057.1 | 840 | 26178.9 | 755 | 23427.6 | 1.1214 | 1.0873 | 1.2411 | 1.1155 | 16–23 | 60–109 |
| 56 | 26517.9 | 755 | 25673.9 | 735 | 29304.2 | 840 | 26379.3 | 755 | 23465.2 | 1.1290 | 1.0937 | 1.2497 | 1.1219 | 16–23 | 60–109 |
| 57 | 26741.4 | 755 | 25864.9 | 735 | 29551.3 | 840 | 26579.7 | 755 | 23502.8 | 1.1365 | 1.1000 | 1.2582 | 1.1282 | 16–23 | 60–109 |
| 58 | 26964.9 | 755 | 26055.9 | 735 | 29798.4 | 840 | 26780.1 | 755 | 23540.4 | 1.1439 | 1.1063 | 1.2668 | 1.1345 | 16–23 | 60–109 |
| 59 | 27188.4 | 755 | 26246.9 | 735 | 30045.5 | 840 | 26980.5 | 755 | 23578.0 | 1.1513 | 1.1126 | 1.2753 | 1.1408 | 16–23 | 60–109 |
| 60 | 28765.3 | 860 | 27980.5 | 840 | 31987.4 | 960 | 28765.3 | 860 | 26543.2 | 1.0835 | 1.0538 | 1.2047 | 1.0835 | 16–23 | 60–109 |
| 61 | 28988.8 | 860 | 28171.5 | 840 | 32234.5 | 960 | 28965.7 | 860 | 26580.8 | 1.0903 | 1.0597 | 1.2128 | 1.0897 | 16–23 | 60–109 |
| 62 | 29212.3 | 860 | 28362.5 | 840 | 32481.6 | 960 | 29166.1 | 860 | 26618.4 | 1.0970 | 1.0655 | 1.2206 | 1.0958 | 16–23 | 60–109 |
| 63 | 29435.8 | 860 | 28553.5 | 840 | 32728.7 | 960 | 29366.5 | 860 | 26656.0 | 1.1038 | 1.0714 | 1.2285 | 1.1018 | 16–23 | 60–109 |
| 64 | 29659.3 | 860 | 28744.5 | 840 | 32975.8 | 960 | 29566.9 | 860 | 26693.6 | 1.1106 | 1.0773 | 1.2363 | 1.1079 | 16–23 | 60–109 |
| 65 | 29882.8 | 860 | 28935.5 | 840 | 33222.9 | 960 | 29767.3 | 860 | 26731.2 | 1.1173 | 1.0831 | 1.2441 | 1.1139 | 16–23 | 60–109 |
| 66 | 30106.3 | 860 | 29126.5 | 840 | 33470.0 | 960 | 29967.7 | 860 | 26768.8 | 1.1241 | 1.0890 | 1.2519 | 1.1200 | 16–23 | 60–109 |
| 67 | 30329.8 | 860 | 29317.5 | 840 | 33717.1 | 960 | 30168.1 | 860 | 26806.4 | 1.1309 | 1.0949 | 1.2597 | 1.1260 | 16–23 | 60–109 |
| 68 | 30553.3 | 860 | 29508.5 | 840 | 33964.2 | 960 | 30368.5 | 860 | 26844.0 | 1.1377 | 1.1007 | 1.2675 | 1.1320 | 16–23 | 60–109 |
| 69 | 30776.8 | 860 | 29699.5 | 840 | 34211.3 | 960 | 30568.9 | 860 | 26881.6 | 1.1444 | 1.1066 | 1.2753 | 1.1380 | 16–23 | 60–109 |
| 70 | 32353.7 | 970 | 31433.1 | 945 | 49938.2 | 1500 | 32191.2 | 965 | 30581.1 | 1.0580 | 1.0278 | 1.6330 | 1.0525 | 16–23 | 60–109 |
| 71 | 32577.2 | 970 | 31624.1 | 945 | 50285.3 | 1500 | 32391.6 | 965 | 30618.7 | 1.0640 | 1.0330 | 1.6424 | 1.0578 | 16–23 | 60–109 |
| 72 | 32800.7 | 970 | 31815.1 | 945 | 50632.4 | 1500 | 32592.0 | 965 | 30656.3 | 1.0699 | 1.0381 | 1.6517 | 1.0632 | 16–23 | 60–109 |
| 73 | 33024.2 | 970 | 32006.1 | 945 | 50979.5 | 1500 | 32792.4 | 965 | 30693.9 | 1.0758 | 1.0433 | 1.6609 | 1.0685 | 16–23 | 60–109 |
| 74 | 33247.7 | 970 | 32197.1 | 945 | 51326.6 | 1500 | 32992.8 | 965 | 30731.5 | 1.0817 | 1.0484 | 1.6701 | 1.0738 | 16–23 | 60–109 |
| 75 | 33471.2 | 970 | 32388.1 | 945 | 51673.7 | 1500 | 33193.2 | 965 | 30769.1 | 1.0876 | 1.0536 | 1.6793 | 1.0791 | 16–23 | 60–109 |
| 76 | 33694.7 | 1080 | 32579.1 | 1111 | 57917.5 | 3000 | 31713.6 | 1111 | 30954.5 | 1.0886 | 1.0526 | 1.8712 | 1.0245 | 16–127 | 60–428 |
| 77 | 33918.2 | 1080 | 32770.1 | 1111 | 61096.2 | 3000 | 31813.1 | 1111 | 31428.4 | 1.0795 | 1.0428 | 1.9440 | 1.0122 | 16–127 | 60–428 |
| 78 | 34141.7 | 1080 | 32961.1 | 1111 | 65763.8 | 3000 | 32229.9 | 1111 | 31687.2 | 1.0775 | 1.0401 | 2.0754 | 1.0171 | 16–127 | 60–428 |
| 79 | 34365.2 | 1080 | 33152.1 | 1111 | 57716.8 | 3000 | 32794.9 | 1111 | 31583.7 | 1.0884 | 1.0494 | 1.8274 | 1.0383 | 16–127 | 60–428 |
| 80 | 35987.5 | 1080 | 34212.5 | 1111 | 67889.0 | 3000 | 35605.4 | 1111 | 34612.9 | 1.0397 | 1.0418 | 1.9614 | 1.0287 | 28–127 | 73–428 |